\documentclass[acmsmall,screen,noacm]{acmart}

\usepackage{graphicx}
\usepackage{subcaption}
\usepackage{multirow}
\usepackage{adjustbox}
\usepackage[font=small,labelfont=bf]{caption}
\renewcommand\footnotetextcopyrightpermission[1]{}
\AtBeginDocument{%
  }

\setcopyright{none}
\renewcommand\footnotetextcopyrightpermission[1]{} 

\begin{document}

\title{Testing Autonomous Driving Systems – What Really Matters and What Doesn't}

\author{Changwen Li}
\affiliation{%
  \institution{Key Laboratory of System Software (CAS), ISCAS}
  \city{Beijing}
  \country{China}}
\email{licw@ios.ac.cn}

\author{Joseph Sifakis}
\affiliation{%
  \institution{University Grenoble Alpes, CNRS, Grenoble INP, VERIMAG}
  \city{Grenoble}
  \country{France}
}

\author{Rongjie Yan}
\affiliation{%
  \institution{Key Laboratory of System Software (CAS), ISCAS}
  \city{Beijing}
  \country{China}}
\email{yrj@ios.ac.cn}

\author{Jian Zhang}
\affiliation{%
  \institution{Key Laboratory of System Software (CAS), ISCAS}
  \city{Beijing}
  \country{China}}
\email{zj@ios.ac.cn}

\begin{abstract}
\textbf{Abstract.} Despite extensive research, the testing of autonomous driving systems (ADS) landscape remains fragmented, and there is currently no basis for an informed technical assessment of the importance and contribution of the current state of the art. This paper attempts to address this problem by exploring two complementary aspects. 

First, it proposes a framework for comparing existing test methods in terms of their intrinsic effectiveness and validity. It shows that many methods do not meet both of these requirements. Either because they are based on criteria that do not allow for rapid, inexpensive, and comprehensive detection of failures, or because the degree of validity of the properties tested cannot be accurately estimated. In particular, it is shown that most critical test methods do not take into account the nominal operational capabilities of autopilots and generate scenarios that are impossible for the tested vehicles to handle, resulting in unjustified rejections.

Secondly, the paper shows that test effectiveness and validity are highly dependent on how autopilots are designed: how they choose between different control policies to perform maneuvers, as well as on the reproducibility of the results. In fact, most test methods take for granted two principles underlying traditional methods, but do not generally apply to ADS. In particular, they implicitly assume that autopilots are rational: passing a critical test implies passing less critical tests, meaning that autopilots are consistent in their choice between different possible control policies. Furthermore, they rely on coverage criteria that assume determinacy of control policies: passing a particular maneuver from a given state implies passing maneuvers from all intermediate states involved in that maneuver. We maintain that the absence of rationality and determinacy significantly impairs the effectiveness and validity of test methods, and provide test results on eight open autopilots, in which most do not satisfy these properties, thereby illustrating this fact.

We conclude that under the current state of the art, it is impossible to obtain strong enough guarantees for essential autopilot properties and recommend that autopilots be developed with a view to both rationality and determinacy.
\end{abstract}

\maketitle

\section{Introduction}

\label{sec:introduction}

ADS testing has been the subject of much work and great interest on the part of the scientific and technical community for two reasons. On the one hand, ADS are extremely complex systems due to their heterogeneity and the intricate dynamics of their environment, involving a large number of interdependent parameters related to road topology, the distribution of moving objects, and applicable traffic rules. On the other hand, strong economic incentives are pushing the search for evidence that existing autopilots can be as safe as human-driven vehicles.

The test methods applied to ADS cover a wide range, from statistical techniques to rule-based test generation techniques or machine learning techniques. Test case generation must take into account various coverage criteria concerning road structure, distribution of moving objects, and operational uncertainties. There are generic methods that use statistical data on traffic patterns to replicate these patterns by selecting appropriate benchmarks. Other methods seek to generate scenarios likely to reveal defects by creating extreme situations based on various risk indicators.

Despite the huge number of works on ADS testing, it must be noted that they remain fragmented and do not provide a unified approach to the problem of ADS validation and evaluation. One important question is the relevance of all these works and what they tell us about the safety of existing solutions. Are the identified risks well-founded, and do they reflect real weaknesses? To what extent do the numerous and varied approaches allow objective and indisputable conclusions to be drawn about the degree of safety of ADS? 

It must be acknowledged that the multitude of existing works seems to be motivated by a certain urgency to reach a conclusion, whatever it may be, without paying sufficient attention to the foundations of testing as a method of producing empirical knowledge. There is currently no basis for an informed and technical assessment of the different approaches and their potential contribution to a sound safety assessment of ADS. This requires rigorous analysis in light of the requirements posed by the epistemic imperatives that any empirical test validation method must satisfy in order to produce valid, reliable, and justifiable knowledge.

Testing is the fundamental approach to producing empirical knowledge in the field of science and technology. Test methodologies must be transparent and clearly state the hypotheses relating to the system being tested and the criteria for their validation. Furthermore, the criteria must be defined objectively, leaving no room for subjective or arbitrary interpretations. Finally, test results must be reproducible to confirm their validity.

For ADS, we distinguish between two main approaches for testing a designated vehicle, generally referred to as the \textit{ego vehicle}. Generic testing, in which the ego vehicle drives on public roads where traffic conditions are uncontrollable and unpredictable. Although generic testing is necessary, it does not effectively cover rare incidents involving complex scenarios. For this reason, critical-scenario-based testing has been proposed to focus on specific scenarios designed to create situations most likely to reveal potential defects. Unlike generic tests carried out in real-world conditions where everything runs normally over relatively long periods of time, critical-scenario-based tests are short and are carried out in simulated environments or closed-course testing environments, which are more accessible, flexible, and scalable. They are generated using different approaches: extraction of scenarios from naturalistic datasets (e.g., Waymo Open Dataset \cite{sun2020scalability}, nuScenes \cite{caesar2020nuscenes}); rule-based generation using predefined logic; ML-generated scenarios used in particular to discover edge cases; and fuzzing techniques for stress testing to randomly disrupt operation and find failure modes. There is a wealth of literature on these different approaches, described in detailed studies \cite{zhang2021finding}. Our objective is to provide a methodological framework for estimating the potential contribution of each approach, its limitations, and to identify gaps that hinder a unified approach for empirical validation and evaluation of autopilot properties.

Applying a test case produces a scenario in the form of a sequence of scenes. A scene is a snapshot of the state of the system, including the kinematic state of moving objects, as well as the dynamic relationships between them. The environment in which vehicles are tested can consist of public roads, closed-course environments, or simulated environments. Public roads offer the least controllability and require the use of generic test methods where the ego vehicle follows a predefined route and surrounding objects are real traffic participants. Closed-course environments offer greater controllability, as the behavior of surrounding objects can be controlled using predefined policies. Finally, simulated environments offer the highest level of controllability, allowing for the testing of highly critical situations, as well as customized maps and weather conditions.

A scenario describes the evolution of ADS in two parts: 
\begin{enumerate}
    \item A static part, which specifies the road environment of the ego vehicle, the road layout, the configurations, as well as the number of moving obstacles involved. It is usually represented in the form of a map containing all relevant information relating to signage.
    \item A dynamic part, which describes the evolution of the system from the initial state of the moving objects on the map, in the form of a sequence of their states.
\end{enumerate}

More specifically, a scenario is generated after applying a test case to validate properties of the ego vehicle. It is a sequence of states $(q_0,s_0),(q_1,s_1),…,(q_n,s_n)$, where $q_i,s_i$ are respectively, the states of the ego vehicle and of the considered mobile objects at the $i$-th step. The state of an object is defined by its position on the map, its speed, and any other relevant attributes.

A test case is a stimulus to the ego vehicle, for which the ego vehicle produces a corresponding response. It is specified by a tuple $(q_0, s_0,...,s_n)$, composed of $q_0$, which is the initial state of the ego vehicle, and a sequence of states of the moving objects. For a given test case, the ego vehicle produces a corresponding control policy, a sequence of states $q_0,..., q_n$. As explained later, the movement of the objects involved in a test case is adequately controlled so as to constrain the movement of the ego vehicle and create situations for which its observed behavior is relevant for the tested properties.

Generated scenarios are checked by an Oracle against safety, performance properties, and even traffic rules expressed as predicates on scenario states. Given a scenario, the Oracle delivers a verdict “pass” or “fail” for each tested property.

\subsection{Key characteristics of an ADS test method}
\label{sec:characteristics}

Defining a test method for ADS involves a series of important choices, which determine the framework within which experiments take place and are discussed below. These include operating conditions, the formulation of the properties to be tested, test coverage, and the evaluation of test results.

\textbf{Operating conditions}: Test case generation should take into account the conditions under which the ego vehicle and its environment operate, considering the choice of test cases and the controllability of their parameters. There are generally two types of operating conditions: nominal and non-nominal conditions.

Nominal conditions refer to the standard environment in which an autonomous vehicle is designed to operate, as well as the nominal safety and performance requirements set by its developers. These conditions are typically specified through an operational design domain (ODD) \cite{thorn2018framework}, which specifies the environment, road, and technical constraints under which the vehicle is intended to operate normally.

Nominal operation for ADS implies that the vehicle, as an electromechanical system, has no defects and faithfully executes the commands of its autopilot. Furthermore, the autopilot does not have to deal with vehicle equipment failures such as flat tires, brake failures, steering failures, or malicious attacks. 

When it comes to the environment, nominal conditions imply compliance with rules and regulations regarding the characteristics of different types of roads: structured roads with clear road markings and their equipment, including signage and traffic lights, so that priority between conflicting vehicles is clearly established; the roads are also free of temporary construction sites. Additionally, nominal conditions make assumptions that all moving objects comply with traffic regulations and roads are free of temporary construction sites.

Finally, for simulation-based testing, nominal operating conditions should be realistic and reflect real-world situations as closely as possible. Sometimes, autopilots have acceleration rates that are too low or too high, or limited speed ranges, which can lead to erroneous conclusions about the ability of the tested autopilots to cope with critical situations.

Unlike nominal conditions, non-nominal operating conditions refer to scenarios in which an ADS operates outside its standard, expected, or intended parameters. These conditions may arise due to external factors, internal failures, or unusual demands, which can affect performance and safety. Their testing requires specific techniques that go beyond standard tests, such as fault injection techniques \cite{tian2025testing} and closed-loop hardware simulation techniques. This work focuses on tests under nominal operating conditions.

\textbf{Test coverage}: Test coverage measures the extent to which the applied test cases cover all aspects of the behavior of the system under test. It is typically estimated using metrics defined on a model representing the possible choices to be exercised by the test cases. For example, common metrics for software systems include instruction, branch, and path coverage. Coverage techniques require that test results be reproducible, which means gradually exploring situations while distinguishing between those that have already been tested and those that remain to be tested. Unfortunately, many autopilots do not meet this requirement, which is expressed by a property of determinacy of their control policies discussed in Section \ref{sec:determinacy}.

The coverage criteria for an ADS involve two independent dimensions. One dimension is the static part, represented by a map and traffic pattern of moving objects, in which the scenario generated by a test case unfolds. The choice of static parts should adequately cover the road layouts and patterns characteristic of different types of traffic configurations. The other dimension concerns the choice of dynamic parts involving behavioral parameter values, such as the definition of the initial configuration on the map of the moving objects and the control policies of objects other than the ego vehicle. 

The definition of rigorous coverage criteria for ADS is a challenging problem given the complexity of the test space. To achieve “adequate coverage”, it is necessary to explore a space-time domain, first in terms of road configuration and initial distribution of moving objects, then in terms of the dynamics of these objects, other than the ego vehicle, throughout the scenarios. Exploration can be guided by varying coverage and criticality criteria.

\textbf{Tested properties and their formulation}: ADS test methods take into account safety and performance properties expressed as conditions between observable and controllable state variables of a scenario evaluated by an Oracle. Differences in the formulation of a property can significantly affect test results. For some test methods, safety is limited to the absence of collisions. Other methods do not use explicitly stated properties, but take into account factors that could lead to dangerous situations, such as indices that measure the risk associated with a scenario. Still others perform a more detailed analysis of safety incidents, taking into account traffic violations as well.

It is clear that the way in which properties are formulated determines the accuracy of the results. In particular, there is a big difference between an actually observed violation of properties and the risk inferred from the values of surrogate risk metrics.

\textbf{Evaluation methods}: Evaluation methods are used to estimate the likelihood that the ego vehicle meets the tested properties based on the results of the experiments. There are two types of methods depending on the criteria adopted for risk assessment.	

Quantitative evaluation methods are applied to generic test methods that typically consider a large number of scenarios. They rely on risk measures that range from very simple, such as the distance traveled without fatal accidents, to more complex ones, expressed as the weighted sum of the number of different types of incidents according to their severity. Quantitative evaluation is usually adopted by the automotive industry, which favors statistical methods involving real or simulated tests for long scenarios. 

On the other hand, qualitative evaluation methods are focused on analyzing the root causes of risky situations and, ultimately, on identifying responsibilities in the event of incidents or traffic violations. These methods are applied to short critical scenarios, as explained below.

\subsection{Effectiveness and Validity of ADS test methods}

We analyze existing approaches to ADS testing, based on two key requirements: effectiveness and validity.

\textbf{Effectiveness}: Test methods must be effective in that they enable defects to be detected quickly, inexpensively, and comprehensively. Additionally, the degree of validity of the property being tested should be estimated accurately. Effectiveness is directly related to test coverage and the detection of critical defects. This is also indirectly related to the degree of validity of the method, as this decreases when defects are omitted or reported incorrectly, as indicated below.

Test cases must be designed to reveal discrepancies between the behavior of the ego vehicle and the properties being tested. In particular, their choice should be guided by the goal of the test method. For example, to test robustness, test cases should produce small perturbations, as the goal is to verify that they have a commensurable effect. Conversely, to test safety, they must target extreme values that lead to situations that can reveal defects.

A common analysis criterion in systems engineering is to focus on critical test cases that place the ego vehicle in difficult-to-manage situations. Critical test cases can be obtained either by analyzing real scenarios leading to dangerous situations \cite{najm2007precrash}, or by basing them on criticality criteria \cite{westhofen2023criticality}. These allow test cases sharing the same road layout and patterns to be classified according to the estimated degree of difficulty in successfully completing the corresponding driving operation. Intuitively, the difficulty of driving the ego vehicle depends on the values of behavioral parameters. It can increase when we change the positions and speeds of moving objects around it in such a way as to restrict its freedom of movement. Rigorously defined criticality criteria imply a partial order among test cases. Using such an order, it is possible to theoretically determine the boundaries between situations of minimal and maximal risk, as well as situations at the limits, sometimes referred to as corner cases.

\textbf{Validity}: In test methods, validity refers to the degree to which their results accurately reflect the phenomena they claim to validate and measure. In particular, test cases must be carefully designed to avoid two critical methodological errors: \textit{false rejection} and \textit{false acceptance}. In ADS testing, this type of error can appear under the following conditions.

False rejection can occur when high-criticality test cases place the ego vehicle in situations that cannot be handled based on its nominal capabilities. This often occurs in test methods that focus solely on creating extreme situations without a preliminary analysis that takes into account the dynamic characteristics of the ego vehicle, such as maximum acceleration and deceleration rates. It is very easy to create unmanageable situations that can lead to accidents for which the ego vehicle is not responsible. For example, some test methods use NPCs to create uncontrollable risky situations \cite{li2025comprehensive}, such as cutting off the road with a sudden maneuver, or aggressively violating traffic rules. Pushing toward critical situations by ignoring constraints that result from ODDs and design limitations of the ego vehicle can significantly skew evaluation results.

Similarly, false acceptance can occur if the acceptance threshold of the Oracle is set too loosely. In particular, in non-critical test cases that accept a large number of correct results, passing a test may simply be a matter of luck. A well-designed test case should challenge the ego vehicle's ability to handle and adapt to a situation. A test case $x$ must create a condition such that the ego vehicle is caused to change its behavior and perform an observable action $y$; if, after applying $x$, we observe $y$, the test is successful, otherwise it fails. A test that allows a large number of acceptable results has limited relevance. It gives the ego vehicle sufficient freedom to succeed without adapting its behavior. Take, for example, a case that presents the ego vehicle with a green light at a distance and speed such that it can pass without changing its speed. Thus, it is not possible to determine whether success is the result of appropriate driving or reckless behavior.

The above analysis implicitly establishes a link between the validity of test methods and the concept of criticality. Care should be taken to select test cases within the maximum and minimum criticality limits. In order to avoid false rejection, the most critical test cases must allow for control policies that are feasible under nominal operating conditions of the ego vehicle. In order to avoid false acceptance, low-criticality test cases must be sufficiently discriminating and include pass criteria that require a significant adaptation of the observed behavior.

Another finding of our analysis is that test effectiveness and validity are highly dependent on how autopilots are designed: how they choose between different control policies to perform maneuvers, as well as on the reproducibility of the results. In fact, most test methods take for granted two principles underlying traditional methods that do not generally apply to ADS. First, they implicitly assume that autopilots are rational: passing a critical test implies passing less critical tests, meaning that autopilots are consistent in their choice between different possible control policies. Second, they rely on coverage criteria that assume determinacy of control policies: passing a particular maneuver from a given state implies passing maneuvers from all intermediate states involved in that maneuver. We maintain that the absence of rationality and determinacy significantly impairs the effectiveness and validity of test methods. We illustrate this fact by providing the results of tests performed on eight open autopilots, most of which fail to satisfy these properties.
The paper is organized as follows.

Section \ref{sec:effectiveness-and-validity} deals with the requirements for effectiveness and validity of the ADS test method, arguing that most existing test methods fall short of these requirements. It proposes a rigorous framework to ensure the effectiveness and validity of tests for four classes of highly critical elementary scenarios that lend themselves to analysis in order to avoid erroneous acceptances.
 
Section \ref{sec:autopilot-requirements} shows how the proposed theoretical framework can be used to highlight the lack of rationality and determinacy in eight open autopilots. It proposes a methodology that, for autopilots that satisfy these properties, greatly simplifies the complexity of testing.

Section \ref{sec:conclusion} concludes with a discussion of the implications of our analysis, arguing that the current state of the art has significant shortcomings that could be overcome if autopilots were developed from the outset with rationality and determinacy in mind.

\section{Effectiveness and validity of ADS test methods}
\label{sec:effectiveness-and-validity}

\subsection{Testing ADS}

\subsubsection{Preliminary definition} We consider an ADS consisting of an ego vehicle in interaction with its environment, involving a set of static or dynamic objects. We denote by $Q$ the set of the states of the ego vehicle and by $S$ the set of the states of its environment. 
\begin{itemize}
    \item A scenario of an ADS is a sequence $(q_0,s_0 ),(q_1,s_1 ),…,(q_n,s_n)$, of successive states where $q_0$ and $s_0$ are the initial states of the ego vehicle and of the environment.  
    \item A goal of the ego vehicle is $G=(Q_T,P)$, where $Q_T \subseteq Q$ is a subset of system target states and $P$ is a set of tested properties. $P$ can be safety or performance properties expressed as predicates on the observed system state sequences. They include collision detection, violation of traffic rules, and also performance indices such as road occupancy. $Q_T$ characterizes desirable final states reached at the end of a scenario.
    \item A scenario $sc$ is successful if all its states $(q_0,s_0),(q_1,s_1 ),…,(q_n,s_n)$ satisfy $P$ and the ego vehicle's final state $q_n$ belongs to $Q_T$. 
\end{itemize}

\begin{figure}[h]
    \centering
    \vspace{-10pt}
    \includegraphics[width=0.7\linewidth]{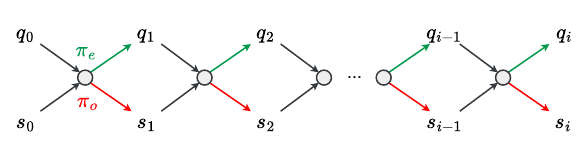}
    \vspace{-20pt}
    \caption{State evolution in a scenario}
    \label{fig:state_seq}
    \vspace{-10pt}
\end{figure}

Note that the sequence of successive states in a scenario can be calculated as part of a highly synchronized interaction involving two actors: the ego vehicle and its environment (Figure \ref{fig:state_seq}). At each step of this interaction, from the current global state $(q_i,s_i)$, the ego vehicle and the objects of the environment decide their next states using two functions $\pi_e$ and $\pi_o$ respectively such that $\pi_e(q_i,s_i)= q_{i+1}$ and $\pi_o( q_i,s_i)= s_{i+1}$. That is, based on the global state, each one of the actors chooses its next state. 

A test case is a stimulus for the tested autopilot. It is denoted by $(q_0,s[n])$, specifying the initial state of the ego vehicle and a sequence of $n$ states $s[n]$ of its environment. In response to this stimulus, the ego vehicle applies a control policy function $pol(q_0,s[n])=q[n]$  that generates a corresponding sequence of states $q[n]$ that is the policy applied by the autopilot. So, the corresponding scenario is $sc=(q[n],s[n])$. 

It is clear that for a given $tc$, there may be different policy functions leading to different control policies. The autopilot of the ego vehicle is supposed to select and apply one of them according to internal criteria seeking adaptation to external stimuli and the satisfaction of essential properties. In some cases, we will consider that the function $pol$ can be computed by sequential applications of a function $\pi_e(q_i,s_i )= q_{i+1}$.

Generating critical scenarios that comply with the nominal conditions of the ego vehicle involves technical difficulties. How can we calculate the reactions of the environment (function $\pi_o$) in such a way as to constrain the behavior of the ego vehicle as much as possible while leaving it a safe margin of reaction for nominal operation? This requires very costly predictive analysis if we want to be accurate in our forecasts. 

One way to overcome these technical difficulties is to consider \textit{adverse scenarios} in which the ego vehicle, in the event of a conflict with other moving objects, has a lower priority. This implies that moving objects of the environment are not required to take the ego vehicle's movement into account, but only to follow the rules of the road. In that case, the function $\pi_o$ can be practically independent of the state of the ego vehicle. Thus, for a given initial state $(q_0,s_0)$, we can generate a sequence of states of the environment $s[n]= (s_0,\dots,s_n)$ that determines its behavior while respecting only the initial free space of the ego vehicle and the traffic signals. In other words, for adverse scenarios, all test case parameters can be assigned independently of the ego vehicle’s behavior and then applied to the tested autopilot, thereby simplifying the scenario generation process.

Finally, note that focusing on adverse scenarios allows us to concentrate on the most critical situations for the ego vehicle, which is the desired objective.

\subsubsection{Characterizing effectiveness and validity of test methods}
We discuss below specific parameters of test methods regarding their effectiveness and validity.

Let $NS$ be the set of scenarios that, according to the nominal operating conditions of the ego vehicle, are safe. These scenarios determine the set of situations that the ego vehicle is expected to handle safely and effectively. In other words, if all the scenarios of $NS$ satisfy the tested properties, then validation is complete.

For a given test method, let $GS$ be the set of potentially generated scenarios for different coverage criteria. For instance, $GS$ may consider cases of driving on highways with different types of driving operations, such as following a front vehicle, overtaking, entering, and exiting a highway. Ideally, we would have $NS\subseteq GS$, that is to have full coverage of all the possible nominal scenarios. But additionally, to avoid false rejection, we should require that $GS\subseteq NS$ because non-nominal scenarios would distort evaluation results. So in practice, a realistic objective is to consider strict inclusion $GS\subset NS$.

Let $CS$ be the set of critical scenarios that can be obtained by instantiating scenarios of $GS$ according to a criticality criterion. Using this criterion, it is possible to simplify the problem by retaining only the most critical scenarios within $GS$ and discarding those that are less critical. In fact, if a criticality order between scenarios is defined, it is possible for a given set of scenarios $S$ to define $cs(S)$, which is the set of the most critical scenarios of $S$. So, ideally $CS$ could be $CS=cs(GS)$. But precisely computing $cs(GS)$ may involve complexities that could be avoided in practice. We can simply require that $cs(GS) \subseteq CS\subseteq GS$. This means that we have a systematic manner to replace low criticality test cases with more critical ones, as discussed later.

\begin{figure}[htbp]
  \centering
  \begin{subfigure}[b]{0.25\textwidth}
    \includegraphics[width=\linewidth]{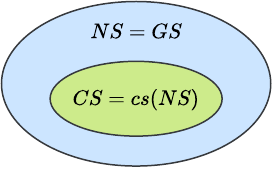}
    \caption{Ideal situation: $NS=GS$ and $CS$ is maximally critical}
    \label{fig:sub1}
  \end{subfigure}
  \hfill
  \begin{subfigure}[b]{0.25\textwidth}
    \includegraphics[width=\linewidth]{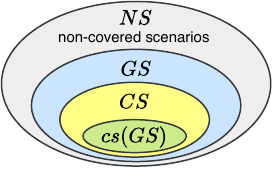}
    \caption{$GS$ for specific scenarios and $CS$ covers $cs(GS)$}
    \label{fig:sub2}
  \end{subfigure}
  \hfill
  \begin{subfigure}[b]{0.305\textwidth}
    \includegraphics[width=\linewidth]{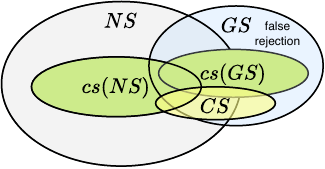}
    \caption{False rejection and $CS$ only partially covers $cs(GS)$}
    \label{fig:sub2}
  \end{subfigure}
  \vspace{-5pt}
  \caption{Possible relations between $NS$, $GS$ and $CS$}
  \label{fig:critical-nominal}
\end{figure}

Figure \ref{fig:critical-nominal} shows possible relations between $NS$, $GS$ and $CS$. The ideal situation corresponds to full coverage and accurate consideration of the corresponding critical scenarios. In practice, we have only partial coverage, $GS\subset NS$, and $CS$ should be a subset of $GS$ containing all the most critical scenarios of $GS$, i.e., $sc(GS)$, while also incorporating a tolerance for including less critical ones. The third case is the one very often observed involving false rejection and also non-systematic coverage with critical scenarios of $GS$.

\subsection{Succinct analysis of the state of the art}
\label{sec:sota}

We present a brief analysis of ADS test methods, focusing on their effectiveness and validity, taking into account key characteristics  such as operating conditions, test coverage, properties tested, and evaluation approaches. The findings are summarized in Table \ref{tab:characterization}.

\begin{table}[h]
    \centering
    \caption{Characterization of the test methods for ADS}
    \label{tab:characterization}
    \vspace{-10pt}
    \adjustbox{max width=\linewidth}{
    \begin{tabular}{|c|c|c|}
    \hline
        \multicolumn{2}{|c|}{\textbf{Characterization}} & \textbf{Options} \\
    \hline
        \multirow{2}{*}{
        \begin{tabular}{c}
             \textbf{Operating}\\
             \textbf{conditions}
        \end{tabular}
        } & \textbf{Test environment} & 
        \begin{tabular}{p{0.9\linewidth}}
            Public Road, Closed course, Simulation
        \end{tabular} \\
        \cline{2-3}
        & \begin{tabular}{c}
             \textbf{Surrounding object}\\
             \textbf{behavior models}
        \end{tabular} & 
        \begin{tabular}{p{0.9\linewidth}}
            $\bullet$ The object is a human participant \\
            $\bullet$ The object is driven by an autopilot \\
            $\bullet$ The object is an NPC with given trajectories (NPC-traj), or driven by controlled policies (NPC-pol)
        \end{tabular} \\
    \hline
    \multirow{3}{*}{
    \begin{tabular}{c}
             \textbf{Test}\\
             \textbf{coverage}
        \end{tabular}
    } & 
    \begin{tabular}{c}
         \textbf{Static part}\\
         \textbf{of scenarios}
    \end{tabular}
    & 
    \begin{tabular}{p{0.9\linewidth}}
            $\bullet$ Random exploration on given map areas \\
            $\bullet$ Data-based method based on given map areas, from accident report, or from real-world recording \\
            $\bullet$ Model-based method by compositional reasoning from an environment model or a combinatorial construction from an ontology-based environment model
        \end{tabular} \\
    \cline{2-3}
    &
    \begin{tabular}{c}
         \textbf{Dynamic part}\\
         \textbf{of scenarios}
    \end{tabular}
    & 
    \begin{tabular}{p{0.9\linewidth}}
            $\bullet$ Increasing the length of scenarios \\
            $\bullet$ Criticality order on initial state of objects by density, worst-case potential scenario, proportion of unsafe possible scenario evolution \\
            $\bullet$ Criticality order on test case by safe policy space, or by size of integral drivable area of the ego vehicle over time\\
            $\bullet$ Criticality order on scenario measured by safety surrogate measures
        \end{tabular} \\
    \cline{2-3}
    &
    \begin{tabular}{c}
         \textbf{Compatibility of} \\
         \textbf{nominal conditions}
    \end{tabular}
    & 
    \begin{tabular}{p{0.9\linewidth}}
            $\bullet$ Existence of safe policy guaranteed by using a reference controller, constraint solver, or mathematical analysis and reasoning \\
            $\bullet$ All vehicles controlled by autopilots starting at zero speed \\
            $\bullet$ The behaviors of surrounding objects follow the real-world distribution \\
            $\bullet$ Existence of safe policies not guaranteed
        \end{tabular} \\
    \hline
    \multirow{2}{*}{
        \begin{tabular}{c}
             \textbf{Tested}\\
             \textbf{properties}
        \end{tabular}
    }
    &
    \begin{tabular}{c}
         \textbf{Types of
         properties}
    \end{tabular}
    & 
    \begin{tabular}{p{0.9\linewidth}}
            Accident avoidance, Traffic rule, Performance
        \end{tabular} \\
    \cline{2-3}
    &
    \begin{tabular}{c}
         \textbf{Types of verdicts}
    \end{tabular}
    & 
    \begin{tabular}{p{0.9\linewidth}}
            Qualitative verdict, Surrogate measures
    \end{tabular} \\
    \hline
    \multicolumn{2}{|c|}{
     \textbf{Evaluation methods}
    } & \begin{tabular}{p{0.9\linewidth}}
        Statistical quantitative evaluation, Qualitative evaluation using test matrix
    \end{tabular} \\
    \hline
    \end{tabular}
    }
\end{table}

\subsubsection{Operating conditions}
\label{sec:operating-conditions}

Operating conditions encompass the test environment and the behavior models of the ego vehicle's surrounding objects. Test environments include public roads, closed courses, and simulation environments. On public roads, the surrounding objects are human participants. In closed courses and simulation environments, these objects can be controlled either by autopilots or by non-autonomous agents, often referred to as non-player characters (NPCs). A scenario defines the routes that vehicles controlled by autopilots must follow autonomously from their initial state. For NPCs, on the other hand, a scenario provides their routes with time or speed constraints to be respected in order to place the ego vehicle in specific critical situations.

NPCs' control policies defined by the scenario fall into two approaches. 

The first approach, NPC-traj, uses predefined trajectories for NPCs whose evolutions follow a function $s_{i+1}=\pi_o(s_i)$, so that their states at each time step are independent of the state of the ego vehicle. This approach is applied by DriveFuzz~\cite{kim2022drivefuzz}, ScenarioFuzz~\cite{wang2024dance}, scenoRITA~\cite{huai2023scenorita}, BehAVExplor~\cite{cheng2023behavexplor}. Note that it is important to ensure that the ego vehicle always has priority over NPCs, making the ego vehicle responsible for any accidents observed. Otherwise, accidents may be due to reckless behavior by NPCs and should not be considered failures of the ego vehicle. As indicated in \cite{huai2023doppelganger}, for a test method with randomly sampled initial states for NPCs following constant-speed trajectories, all 1,109 accident scenarios generated were attributed to NPCs. Similarly, \cite{lin2024tm} reports that among the 31 accidents generated by DriveFuzz, 10 (32.3\%) were attributable to NPCs.

The second approach, NPC-pol, uses NPCs whose functions $s_{i+1}=\pi_o(q_i,s_i)$ depend on the state of the ego vehicle $q_i$. It generally allows situations in which the priorities between the ego vehicle and NPCs change dynamically. For example, an NPC changes from high priority to low priority when driving on a main road and an autonomous vehicle cuts it off, or from low priority to high priority when it is the vehicle cutting off the road.

This approach is applied by AV-Fuzzer~\cite{li2020av}, AutoFuzz \cite{zhong2022neural}, and Adversarial Safety-critical Scenario Generation \cite{hao2023adversarial}. However, it is very difficult to ensure that NPC control policies create critical situations for the ego vehicle, which take into account its nominal operational capabilities. In fact, designing control policies for NPCs that strike the right balance between the criticality and feasibility of safety policies for the ego vehicle is as complex as designing a safe autopilot.

A manual analysis of accidents generated by AV-Fuzzer \cite{lu2024diavio} shows that of the 192 accidents reported, 151 (78.6\%) were attributable to NPCs. Furthermore, \cite{yang2025authsim} indicates that generating control policies for NPCs using reinforcement learning involves a proportion of accidents that cannot be attributed to the ego vehicle varying between 5.9\% and 36.7\%, with a total number of accidents ranging from 256 to 290.

In conclusion, both NPC-traj and NPC-pol policies must be carefully designed to avoid false rejections. Specifically, NPC-traj should be used only in adverse test cases, while NPC-pol raises issues of complexity that are difficult to overcome. To address these issues, in Section \ref{sec:optimal-criticality}, we present a method for analyzing elementary adverse scenarios and identifying the most critical NPC-traj policies.

\subsubsection{Test coverage}
\label{sec:test-coverage}

As explained in Section \ref{sec:characteristics}, the test coverage comprises two independent parts: a static part that refers to the road layout and traffic patterns, and a dynamic part that concerns the dynamic evolution of moving objects of interest, characterized by their kinematic state parameters. Therefore, a systematic approach would first consider scenario classes sharing the same static part, $GS_1,\dots, GS_n$, all contained in the set $NS$ under consideration, such that $GS_1\cup GS_2\cup \dots \cup GS_n \subseteq NS$. This means that the test method first determines the $GS_i$s in order to obtain adequate coverage of specific static road configurations corresponding to the static parts of the scenarios. Then, for each $GS_i$, the test method must properly select $CS_i$ such that $cs(GS_i)\subseteq CS_i\subseteq GS_i$.

\paragraph{Coverage of static parts}
To cover the static part of a scenario, generic test methods randomly select the routes for the ego vehicle and the surrounding objects in a given map area~\cite{carla2020leaderboard}. While test coverage increases with the number of generated test cases, covering rare static configurations on the map often requires a very large number of cases. For instance, in the Carla simulator, only two roundabouts exist across the ten built-in maps~\cite{dosovitskiy2017carla}. It is clearly unlikely that sufficient coverage of roundabouts will be achieved by selecting routes at random.

In order to systematically address the coverage of static configurations, two fundamental approaches, data-based and model-based, are adopted in research works.

Data-based methods take a bottom-up approach that identifies common patterns from a given data set. Depending on the type of data source, different extraction and analysis strategies are employed. For HD maps covering large regions, the strategies typically involve identifying road components such as junctions and classifying them according to topological patterns \cite{tang2021systematic}. When using accident reports, the objective is to extract pre-crash patterns that capture the behaviors of objects causing the accidents. This approach has been demonstrated in studies by NHTSA \cite{najm2007precrash}. For real-world driving recordings, such as dashcam videos, AI-based techniques are applied to extract semantic attributes, including road layouts and object moving patterns~\cite{zhang2023building}.

In contrast, model-based methods take a top-down approach that derives a finite set of static configurations from a semantic model of the environment. These methods can rely on compositional reasoning~\cite{li2024rigorous}, assuming that any static configuration can be constructed from a finite set of basic static configurations. It is also possible to use an ontology-based environment model, from which static configurations are derived by instantiating and combining concepts under nominal constraints by means of combinatorial testing~\cite{li2020ontology, li2022comopt}.

\paragraph{Coverage of dynamic parts}
To cover the dynamic part of a scenario, the considerable number of possible system evolutions makes exhaustive exploration infeasible.

Generic test methods are based on the assumption that the number of critical situations increases with the distance traveled by the ego vehicle \cite{kalra2016driving}, while surrounding objects, whether humans or autonomous vehicles, follow a random route on the map. However, these methods do not effectively cover the extreme diversity of possible critical situations. As shown in \cite{kalra2016driving}, demonstrating the safety of an autonomous vehicle at a 95\% significance level and 80\% power level requires over 11 billion miles of driving.

To avoid the limitations of generic testing, most research focuses on critical scenario testing using a criticality order to reduce the scope of exploration, concentrating on the set of critical scenarios $CS$, which is much smaller than $GS$. This order must represent a concept of criticality such that the validation of critical cases must, with a high degree of probability, imply validation for less critical cases. Furthermore, critical cases must not be impossible to handle by the ego vehicle according to its nominal operational capabilities.

Existing works define different types of criticality orders between initial states $(q_0,s_0)$, test cases $(q_0,s[n])$, or scenarios $(q[n],s[n])$.

The criticality order over initial states $(q_0,s_0)$ is considered when both the ego vehicle and surrounding objects are controlled by their own autopilots. It is adopted by many works \cite{li2022comopt, carla2020leaderboard, xia2018test} that use the simulator’s built-in autopilots to control the surrounding objects.

A simple approach to define a criticality order over initial states $(q_0,s_0)$ is to consider the object density, typically measured by the average distance between the ego vehicle and the surrounding objects, or the distances between the two successive objects along a lane, as applied by \cite{gold2016taking,wang2018traffic,xia2018test}. However, solely increasing density does not necessarily increase the possibility of property violations, as high density may also cause vehicles to move slowly, decreasing the potential risks.

Other criticality orders on initial states take into account the evolution of objects' states from their initial state. More specifically, they consider the criticality of an initial state by analyzing the entire set of generated scenarios $SC = \{(q[n],s[n]) \mid from (q_0,s_0)\}$ where the evolution of each object is constrained by its dynamic model. In accordance with this consideration, two main types of criticality orders are defined:

\begin{enumerate}
    \item Orders defined by values of worst-case surrogate measures over scenarios, such as the minimum or integrated value of time-to-collision (TTC) or the distance between the ego vehicle and other objects~\cite{westhofen2023criticality}. Formally, if $sm(sc)$ denotes a surrogate measure of a scenario $sc$, then the criticality of the initial state $(q_0,s_0)$ is defined as: $\min_{sc\in SC} sm(sc)$, characterizing the most unsafe situation that could occur within $SC$.
    \item Orders defined by the proportion of safe scenarios among the generated scenarios from an initial state. Thus, the degree of criticality is quantified as  $\left|\left\{sc\in SC \mid safe(sc)\right\}\right|/|SC|$, where $safe(sc)$ expresses the fact that $sc$ meets the safety properties under consideration~\cite{stumper2018towards}.
\end{enumerate}

However, in both cases, due to the high complexity of analyzing all possible scenarios, these measures are typically estimated using Monte Carlo sampling, which approximates the scenario space by randomly selecting actions at each time step to generate a finite set of trajectories. This method may have limitations in terms of efficiency, particularly when sampling over long time sequences. As an alternative, probabilistic and reachability-based reasoning can be employed on the ADS dynamic model. But this approach often involves simplifying assumptions and approximations that may compromise accuracy. 

In summary, while orders based on initial states offer valid estimations, they are often computationally intensive and potentially imprecise. In addition, when increasing the criticality of initial states, most works do not consider the nominal conditions for the ego vehicle. As shown in \cite{li2025comprehensive}, even with randomly selected initial states for the ego vehicle and surrounding objects generated by the Carla Leaderboard, among the 104 detected accidents, 25 (24.0\%) of them cannot be attributed to the ego vehicle.

The second class of criticality orders, to address the above limitations, considers orders on test cases $tc=(q_0,s[n])$. This allows reasoning about the possible policies applied by the ego vehicle for a given behavior of its environment. The criticality of two test cases can be compared as an inclusion relationship between the respective sets of control policies \cite{li2024rigorous, karimi2022automatic}, or between the integral drivable area over time for the ego vehicle in a given static configuration~\cite{sontges2017computing}.  

As test cases specify the evolution of the environment, it becomes possible to assess whether nominal conditions hold as criticality increases. Some approaches compute safe policies for a given test case using a reference controller \cite{djoudi2020simulation} or a constraint solver~\cite{karimi2022automatic}. This enables the identification of maximally critical test cases by iteratively increasing criticality up to the point beyond which safe control policies cannot be found. Methods for increasing the criticality of a scenario involve adding NPCs to an initially safe scenario, to create new situations in which the initial policy of the ego vehicle is no longer safe because its feasible policy space is reduced~\cite{karimi2022automatic}. Another method considers constant speed movement of surrounding objects and combines quadratic programming to minimize the integral drivable area over time for the ego vehicle, as well as refining the initial states of surrounding objects using binary search~\cite{althoff2018automatic}. However, computing a safe policy for the ego vehicle under a test case can be as complex as designing an autopilot. Using such methods to determine the feasibility of safe policies under nominal conditions may miss critical nominal cases that lie beyond the capabilities of the reference controllers or the solvers. To remedy this, \cite{li2024rigorous} proposes a rigorous framework that links criticality with nominal conditions, enabling the identification of the most critical test case under nominal conditions through mathematical analysis.

The third class of criticality orders includes orders defined based on surrogate safety measures for scenarios $sc$, denoted $sm(sc) = sm(q[n],s[n])$, which depend on the ego vehicle’s control policy $q[n]$. Therefore, the test method shall perform a predictive analysis of the ego vehicle's behavior, based on parameters that the test generation can control, including the initial states of objects, NPC trajectories, or NPC control policies, as described below.

The predictive analysis begins with a dataset consisting of data points in the form $((\theta_1,\theta_2,…,\theta_n ),$ $sm(sc))$, where $\theta_1,\theta_2,…,\theta_n$ are values of the controlled parameters and $sc$ is the corresponding scenario obtained through simulation. Then, the scenario generation method increases the criticality based on the information from the dataset. Depending on the scale of the parameters, either search-based or learning-based methods are applied.

Search-based methods are applied for small-scale controlled parameters, such as initial states or discrete representations of vehicle policies or trajectories. The search uses heuristic or metaheuristic strategies. For example, a genetic algorithm performs crossover between two highly critical data points, denoted as $((\theta_1, \theta_2, …, \theta_n), sm(sc))$ and $((\theta'_1, \theta'_2, …, \theta'_n), sm(sc’))$, where both $sm(sc)$ and $sm(sc')$ are low. The crossovered scenario has parameters $(\theta''_1, …, \theta''_n)$, where each $\theta''_i$ is selected from ${\theta_i, \theta’_i}$. The assumption is that combining parameters from highly critical scenarios is likely to yield a scenario $sc''$ with a lower safety measures such that $sm(sc'') < sm(sm)$ and $sm(sc'') < sm(sc')$. However, this method is only a heuristic and does not guarantee the production of scenarios with the lowest safety measure.

When the parameter space is high-dimensional, involving timed sequences of NPC trajectories or policy functions, learning-based methods, especially reinforcement learning, are employed. These methods use neural networks to model the relationship between the controlled parameters $\theta_1,…,\theta_n$ and the surrogate measure $sm(sc)$, enabling prediction of values of controlled parameters that produce unsafe scenarios with low $sm(sc)$.

It should be noted that both search-based and learning-based approaches are inherently approximate and cannot guarantee identification of the most critical scenarios. Moreover, they cannot ensure that nominal conditions are preserved while increasing the criticality, as surrogate measures only reflect the resulting risk. In most existing works, the ranges for controlled parameters are selected manually in an ad hoc manner, without checking their compatibility with nominal conditions. To fully ensure compatibility with nominal conditions, DoppelTest considers only vehicles controlled by the autopilot under test that start from zero speed \cite{huai2023doppelganger}. However, this method is only a heuristic and does not guarantee the production of scenarios with the lowest safety measures.

In conclusion, most existing approaches, whether they take into account criticality orders on initial states, test cases, or scenarios, fail to strike the right balance between criticality and compliance with nominal operating conditions, generating scenarios that are either too critical but not nominal or nominal but insufficiently critical. As we will discuss in Section \ref{sec:optimal-criticality}, an essential requirement for test methods is a precise characterization of the boundary between criticality and nominal conditions. This enables test case generation to jointly account for both aspects, ensuring that the most relevant scenarios are identified.

\subsubsection{Tested properties}
\label{sec:tested-properties}

For the tested properties, the main areas of interest are safety, performance, and compliance with traffic rules. Each property can be formulated either directly based on observable states from the test environment to yield a qualitative verdict indicating whether the property is satisfied, or formulated indirectly through surrogate measures that are correlated with or serve as abstractions of the tested properties.

Safety properties focus on accident avoidance, which may be formulated as verdicts indicating the occurrence of collisions and relevant context, or using surrogate measures such as Time-to-Collision (TTC) – How much time remains before a potential collision, Deceleration Rate to Avoid Crash (DRAC) – The required braking to prevent a crash, Post-Encroachment Time (PET) – The time difference between two road users passing a conflict point, Conflict Severity – Based on speed, impact angle, and object mass, and Interaction Complexity – Multi-agent interactions (e.g., pedestrians, cyclists, other vehicles). 

Traffic rule compliance is typically validated via programmatic checkers~\cite{carla2020leaderboard} or runtime verification techniques based on temporal logic specifications \cite{li2023simulation, sun2022lawbreaker}. However, it should be noted that most test methods either ignore traffic rule checking or consider only simple rules, such as stopping at red lights or stop signs, as in~\cite{carla2020leaderboard}. Validating compliance with more complex rules, such as priority or mutual exclusion rules at junctions, requires semantic awareness of the scenario, in particular, the semantics of the map and the states of moving objects on the map. Such semantic information is often inaccessible through simulator APIs like those of Carla~\cite{dosovitskiy2017carla} and LGSVL~\cite{rong2020lgsvl}, necessitating additional effort to extend the simulation environment and construct appropriate oracles.

Performance properties can be validated by comparing the observed control policy with a theoretically optimal policy~\cite{li2024rigorous}, or by using measures such as total route completion time~\cite{carla2020leaderboard,li2022comopt}. Although various methods exist for validating performance properties, some test methods focus solely on safety, such as~\cite{li2020av,tian2022mosat}. In fact, safety properties can often be easily satisfied through performance degradation, such as by simply not moving forward. Overlooking performance considerations may therefore hide failures of the tested autopilot.

\subsubsection{Evaluation methods}
\label{sec:evaluation-methods}

Evaluation methods are generally classified into two categories: quantitative and qualitative. 

Among quantitative methods, statistical approaches are widely used. They allow failure rate estimation, accompanied by a significance level and a power level, which characterize the probabilities of correct rejection or correct acceptance of the evaluation result, respectively. When multiple types of failures are considered, the failure rate is a weighted sum of all failures, with each failure rate being assigned a weight based on its severity.

A simple statistical evaluation method from generic test results is to estimate the failure rate of the ego vehicle directly. Alternatively, evaluation can focus on specific high-risk situations, such as merges or intersections. In these cases, Monte Carlo methods can be employed to estimate failure rates by sampling test cases from real-world distributions. However, as discussed in Section \ref{sec:test-coverage}, these approaches require a large number of scenarios to ensure fair coverage of real-life situations in order to produce statistically meaningful results.

To improve the efficiency of statistical techniques, two acceleration methods, importance sampling and subset simulation, can be adopted. 

Importance sampling \cite{feng2023dense} allows test cases to be generated from distributions different from the real-world ones, and estimates the true failure rate by re-weighting outcomes according to the ratio of the real-world and sampling distributions. This technique is able to reduce the number of required test cases when failures are rare events in the real-world distribution.

Subset simulation, on the other hand, decomposes a rare-event probability into a product of sub-events that may occur more frequently. This transformation converts a difficult rare-event estimation into a sequence of easier estimation problems, thereby improving sampling efficiency, as demonstrated in~\cite{zhang2018accelerated}.

However, some works adopt evaluation metrics such as failure rate or weighted failure scores without reporting the corresponding statistical or other indicators of validity to justify that the evaluation results reflect real-world outcomes. Omitting these validity indicators can lead to biased or misleading conclusions. For example, the Carla Leaderboard aggregates the number of failures in non-critical, critical, and non-nominal test cases without explaining how the number of different types of applied test cases relates to the real-world frequencies. A detailed analysis conducted by~\cite{li2025comprehensive} reveals that the Carla Leaderboard evaluation method takes into account many failures that cannot be attributed to the ego vehicle, while failing to detect certain types of accidents that may occur in the real world. In other words, this evaluation method is prone to false acceptance and false rejection errors.

Another type of biased evaluation occurs when failures are counted without taking into account possible similarity or redundancy between test scenarios. As reported in \cite{li2024rigorous}, among the 330 scenarios generated by DoppelTest, 52 resulted in accidents. However, only 4 of these scenarios were distinct, while the other 48 were generated by the 4 logically equivalent test cases, differing only in objects irrelevant to the ego vehicle's driving policy.

In addition to statistical methods, it is also possible to use worst-case scenario evaluation methods \cite{nilsson2015worst}. These methods identify the scenario with the worst surrogate measures for the tested properties among all generated scenarios. However, as noted in Section \ref{sec:test-coverage}, many existing works do not guarantee that the generated test cases meet the nominal conditions or limit test generation to suboptimal criticality levels. As a result, the worst-case scenarios identified may not accurately reflect real-world outcomes.

Qualitative evaluation encompasses various techniques that can be applied independently. Their results are presented in the form of a test matrix organized in a table where each test case is linked to a specific driving situation \cite{li2024rigorous, thorn2018framework}. A test matrix provides detailed information for each individual test case and its outcome. However, most works adopting this approach fail to clarify essential aspects of the evaluation, such as the criticality of the test cases, the nominal conditions, and coverage criteria.

More recently, Large Language Models (LLMs) have been used to assist in assigning responsibility in the event of violations of safety or performance properties \cite{lu2024diavio}, thereby improving the validity of the evaluation for the tested autopilot. Although promising, this approach can produce incorrect evaluation results that differ from human judgment \cite{lu2024diavio}, leading to false acceptances or rejections.

To complete our analysis, we would like to highlight the existence of standards such as ISO/PAS 21448 (SOTIF) and UL 4600, which define conceptual frameworks aimed at improving the validity of safety assessments and testing processes as a whole. These include aspects that fall outside the scope of this work, namely the analysis of failure modes and the identification of risk factors and their mitigation using techniques such as FMEA, HAZOP, and STPA, as described in detail in \cite{sun2022comparison}.

\subsubsection{Conclusion of the analysis}

The analysis of the state of the art reveals numerous shortcomings in test methods in terms of effectiveness and validity. These result from the transposition of traditional test methods without taking into account the specific characteristics of ADS, which, unlike software and hardware systems, exhibit complex dynamic adaptive behavior and whose coverage criteria depend on a large number of factors. While traditional methods select test cases defined in predefined value domains, test case selection for ADS must be performed dynamically, constantly taking into account their operating conditions. Ignoring these conditions, as is often the case, can compromise the validity of the methods. 

We have demonstrated that using NPCs to create critical situations by pushing the constraints applied to the ego vehicle to extremes without taking into account its nominal capabilities to handle these situations, leads to highly erroneous results. Given that dynamically taking these constraints into account remains an extremely complex problem, the only solution is to consider short adverse scenarios in which the ego vehicle does not have priority over vehicles in its environment. In this case, it is sufficient for NPCs to comply with traffic rules.

With regard to test coverage, while not denying the value of generic test methods for validating common situations, these are not sufficient to detect critical situations. Our analysis of different criticality criteria that guide scenario generation shows that, in addition to problems related to neglecting nominal operating conditions, the use of surrogate criteria with heuristic or learning-based methods does not provide the accuracy needed to reproduce failure situations.

Furthermore, the test methods analyzed rarely take into account the properties of the autopilot being tested. As we will see in more detail in Section \ref{sec:autopilot-requirements}, the relevance of coverage criteria depends heavily on these properties. The implicit assumption that success in difficult cases implies success in easier cases is often not satisfied. Furthermore, the concept of coverage assumes the reproducibility of observed behaviors, which is far from guaranteed for current autopilots. Without these considerations, it is impossible to establish valid coverage criteria.

The choice of properties tested largely determines the degree of validity of the test results. A distinction must be made between what is possible and what actually happens, taking into account traffic violations and considering both safety and performance aspects. Analyzing test outcomes without accounting for all relevant properties renders the test results meaningless.

All these limitations are reflected in the difficulty of estimating what is actually being tested and to what extent. They suggest that the validity and reliability of most of the evaluation criteria are questionable in several respects. This is particularly true for quantitative evaluation, which must be based on sound statistical methods that take into account the frequency and severity of incidents according to a typology of the observed situations.

%\subsection{Reliability of ADS test methods}
\subsection{A rigorous framework to ensure effectiveness and validity}
\label{sec:rigorous-framework}

In this section, we propose a rigorous testing framework for the qualitative evaluation of ADS. To avoid the shortcomings identified, the framework focuses on elementary adverse scenarios for which false acceptances and false rejections of test cases can be avoided through simple model-based analysis. The presentation of this framework also allows for a more technical comparison with the main critical-scenario-based methods.

\subsubsection{Avoiding false acceptance}

The purpose of a test is to trigger a reaction of the ego vehicle that produces a scenario for which a verdict is delivered by an Oracle. 

Failure for a scenario means that the tested properties $P$ are not satisfied by some of its states. However, success must demonstrate that the ego vehicle's control policy has been adapted to overcome a difficulty and must be measured using relevant criteria. We will refer to \textit{false acceptance} if these conditions are not met. 

We can imagine conflict-free scenarios in which the ego vehicle follows a vehicle ahead of it on a highway and drives safely by following the same simple collision avoidance policy. Under these conditions, it can travel long distances without incident, which can be considered a success according to some evaluation criteria that assign scores based on distance traveled.

This argument shows that evaluation results based on the number of incidents per distance or travel time, published by industry to prove the safety of their autonomous cars \cite{etherington2019waymo}, may be superficial. What matters is the number of conflict situations encountered and the corresponding outcomes.

However, false acceptance can also occur in adverse scenarios in which it is not possible to establish a cause-and-effect relationship between the test case and the observed result.

A scenario is considered as \textit{irrelevant} if it does not require any modification of the initial speed of the ego vehicle to satisfy the tested properties. 

\begin{figure}[h]
    \centering
    \includegraphics[width=0.4\linewidth]{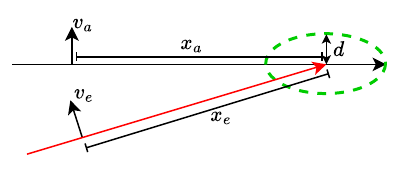}
    \vspace{-10pt}
    \caption{Merge of the ego vehicle into a main road}
    \vspace{-10pt}
    \label{fig:merge-example}
\end{figure}

Consider for instance, a simple scenario involving the ego vehicle and another arriving vehicle traveling on a high-priority road. The route of the ego vehicle merges into the high-priority road as indicated in Figure \ref{fig:merge-example}. 

The initial state of the ego vehicle is characterized by its distance $x_e$ from the merge point and its speed $v_e$, while the arriving vehicle is initially at distance $x_a$ and keeps moving at constant speed $v_a$. Assume that the ego vehicle applies a policy that keeps its speed constant at $v_e$. The generated scenario may fail if a collision occurs. This can happen if the two vehicles are simultaneously in a critical zone defined by road segments within a distance $d$ before or after the merge point. However, there is a high probability that the scenario avoids collision and thus passes the test successfully. In fact, the arriving vehicle can be in the critical zone in times in the interval $[x_a/v_a -d/v_a, x_a/v_a +d/v_a]$ and the ego vehicle can be in the critical zone in times $[x_e/v_e -d/v_e, x_e/v_e +d/v_e]$. It is easy to see that a collision between the two vehicles can occur only when the absolute value $|x_e/v_e - x_a/v_a|\leq d/v_e +d/v_a$. When the speeds and distances do not satisfy this constraint, there is no collision. This is highly likely to occur when the distances between the vehicles and the merge point are sufficiently large. 

If we consider a simple case where $v_e=v_a$, the condition gives $|x_e-x_a |\leq 2d$. For instance, if $d=5$m then for all combinations of values of $x_e,x_a$ where $10<|x_e-x_a|$, collision will not occur despite careless driving of the ego vehicle. Such situations can often occur during generic tests where conflicts between vehicle trajectories are random. Thus, success can falsely count positively when it is purely and simply the result of chance. 	

It is therefore important, for efficiency reasons and to avoid distorting evaluation results, that the test cases place the ego vehicle in a situation such that if it does not modify its speed, it will fail. The modification of the speed is an acknowledgment of the fact that the ego vehicle is sensitive to the external stimulus that constitutes a test case.

Finally, note that success in safety does not imply correctness: the autopilot may pass by being overly cautious, e.g., by stopping unnecessarily, which is not functionally correct. For example, an autopilot that always yields at an empty intersection may pass safety tests but frustrate users and disrupt traffic flow. This is an argument in favor of methods that allow for a comprehensive analysis of all relevant aspects of ADS behavior.

\subsubsection{Avoiding false rejection}

\textit{False rejection} is the opposite of false acceptance, which occurs when the situations created are too difficult or their relevance is overestimated by the evaluation criteria. It occurs when a scenario can fail because the situations created are impossible to handle safely under the vehicle's nominal operating conditions. 	

These conditions must be taken into account in the scenarios, as they determine the behavior generated by the ego vehicle control policies. 

For example, if the ego vehicle is driving on a high-priority road and crosses intersections protected by stop signs, it will assume that vehicles approaching from secondary roads will obey these signs. If a vehicle unexpectedly runs a stop sign and causes an accident, this should not be considered a failure of the ego vehicle's autopilot. Allowing vehicles in the environment to violate traffic rules under unspecified conditions can create unmanageable critical situations. However, most critical-scenario-based test methods ignore this requirement to restrict the behavior of the ego vehicle within its nominal operating conditions. This is particularly true for test methods that use NPCs, as explained above.

Note that the need to avoid false rejections conflicts with the requirement for critical-scenario-based testing to push the system under test to its limits by exposing it to the most challenging situations. Therefore, scenario generation must seek the best possible compromise between these two requirements. The search for such a compromise raises interesting questions about the possibility of generating critical test cases that meet this need.

The various methods for generating critical scenarios and their ability to take into account nominal operating conditions were analyzed in Section \ref{sec:sota}. They adopt an explicit or implicit criticality order constructed from the same empirical observation. Given two scenarios that share the same static part, it is easy to create dynamic parts where the positions of the objects involved are topologically similar but differ in terms of criticality. This can be achieved by modifying characteristics of the dynamic parts that affect the difficulty of applying the corresponding control policies. For example, we can increase the criticality of a scenario by decreasing the distance of obstacles surrounding the ego vehicle or by decreasing the speed of preceding vehicles and increasing the speed of the vehicle with which an unfavorable conflict is possible. Other criticality criteria may include the computational complexity involved in calculating control policies or risk indices depending on the kinematic states of the vehicles under consideration, as already explained.  

\subsubsection{Characterizing the optimal criticality order}
\label{sec:optimal-criticality}

An important question is how to compare the different criticality orders in terms of effectiveness. Furthermore, among all these criticality orders, is there an \textit{optimal order} that structures all nominal test cases according to a reasonable difficulty criterion? We examine this question below, focusing on adverse scenarios linked by a cause-and-effect relationship with the test cases.

We define the criticality order $\prec$ on test cases in the following manner. For two test cases $tc=(q_0,s[n]),tc'=(q_0,s'[n])$, with the same static part, $tc'\prec tc$, if the set of the safe policies for $tc$ is contained in the set of the safe control policies for $tc'$, that is $\{q[n] \mid \exists pol(q_0,s[n])=q[n] \ safe\} \subseteq \{q[n] \mid \exists pol(q_0,s'[n])=q[n] \ safe\}$. 

The application of this definition relies on the hypothesis that for a given test case $tc=(q_0,s[n])$ and context, it is possible to find increasingly critical test cases for which the number of safe policies decreases. One can imagine that this could be achieved by tightening the constraints imposed on the ego vehicle’s operations while keeping the static part of the scenario unchanged, for example, by appropriately modifying the speeds and distances of the surrounding obstacles. 

An interesting property of this definition is that as criticality increases, the number of safe policies decreases. So, if we consider a chain of test cases  $tc_1\prec tc_2 \prec tc_i  \prec \dots$, we can find a maximally critical test case for which the number of safe policies is zero. In addition, for a given test case, there exists a maximal set of safe policies respecting the nominal operating conditions.

\begin{figure}
    \centering
    \includegraphics[width=0.5\linewidth]{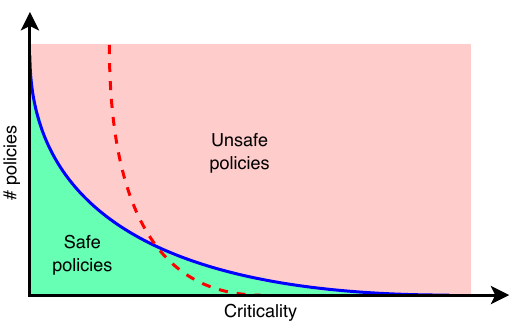}
    \vspace{-5pt}
    \caption{Separation between safe and unsafe policies using the optimal criticality order}
    \label{fig:policy-separation}
\end{figure}

The set of maximal policies for a given $q$ defines a curve in a multidimensional space relating, for each test case, the number of safe control policies. An abstract representation of this curve for a total order between test cases and the number of policies is represented in blue in Figure \ref{fig:policy-separation}. This curve is the set of the limit points beyond which there is no safe control policy for nominal conditions. It corresponds to the best compromise between criticality and safety. Hence, any other criticality order either restricts the set of considered critical test cases or allows test cases leading to unsafe policies, as illustrated by the red curve in Figure \ref{fig:policy-separation}.

This definition of criticality, based on the inclusion of safety policies, can be used to better understand and analyze the behavior of autopilots. An autopilot faced with an unfavorable conflict situation for which there is a non-empty set of safe policies, can make the following choices: 1) Choose an unsafe policy, outside of the set of the safe policies, which may cause failures; 2) Choose a safe policy that neglects performance aspects; or 3) Choose a policy that is safe and also best meets performance requirements. Of course, between 2) and 3), there are intermediate situations that depend on the degree of optimality of the policy chosen in relation to performance requirements. As explained above, these distinctions are important because safety means not making any mistakes, which includes ignoring the possibility of moving when it is possible to do so.

An interesting question is whether we can characterize the optimal criticality order. 

We explained that focusing on adverse test cases facilitates their generation by decoupling the interdependencies between the ego vehicle and its environment, thus making explicit the cause-and-effect relation in testing. We show that, thanks to such a decoupling, it is possible to compute the most critical test cases that are also safe, for \textit{elementary adverse scenarios}, i.e., adverse scenarios involving a single conflict between the ego vehicle and another vehicle arriving from a road that meets its route.

\begin{figure}[htbp]
  \centering
  \begin{subfigure}[b]{0.45\textwidth}  % 宽度可根据需要调整
    \includegraphics[width=\linewidth]{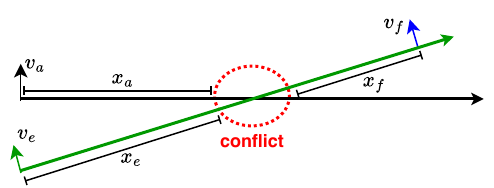}
    \caption{Intersection}
    \label{fig:sub1}
  \end{subfigure}
  \hfill
  \begin{subfigure}[b]{0.45\textwidth}
    \includegraphics[width=\linewidth]{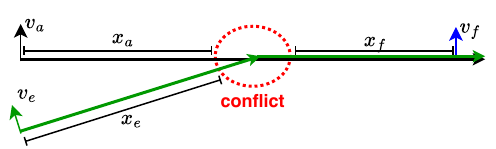}
    \caption{Merge}
    \label{fig:sub2}
  \end{subfigure}
  \vspace{-5pt}
  \caption{The two elementary adverse scenarios}
  \label{fig:ele-adv-scenarios}
\end{figure}

We consider two types of elementary adverse scenarios, whose static parts are shown in Figure~\ref{fig:ele-adv-scenarios}. In the first type, the ego vehicle and the arriving vehicle take different directions after passing the conflict point, forming an intersection scenario. In the second type, both the ego vehicle and the arriving vehicle follow the same route after the conflict point, resulting in a merge scenario. In both scenarios, the ego vehicle should take into account the presence of a front vehicle ahead on its route.

The intersection scenario covers both intersections protected by traffic lights and those protected by yield signs. The merge scenario covers merging into a high-priority road protected by a yield sign, as well as lane changes during an overtaking maneuver. 

In all these scenarios, the ego vehicle is in an unfavorable conflict and seeks to cross a critical zone protected by access rules. The scenarios end when the ego vehicle leaves this critical zone and continues on its way without colliding with the vehicle in front. We compute below critical test cases, where the ego vehicle has no choice but to modify its initial speed at the risk of breaking a safety rule or being less efficient in the case of overtaking. 

Let $v_e,v_a, v_f$ respectively denote the speeds and $x_e,x_a, x_f$ the distances of the ego, the arriving, and the front vehicle from the conflict point as indicated in Figure \ref{fig:ele-adv-scenarios}. We consider test cases of the form $((x_{e_0},v_{e_0}),(X_a[n],V_a[n]),(X_f[n],V_f[n]) )$, where the first element is the initial state of the ego vehicle and the second and third element describe the evolution of the arriving and front vehicle using timed sequences with length $n$. That is, $X_a[n], X_f [n]$ are respectively decreasing and increasing sequences of values giving the distance at each time step, and $V_a[n]$, $V_f[n]$ are the speed sequences. We assume that the speed of each vehicle is bounded by the speed limits for each road. In particular, let $vl$ be the speed limit applied to the arriving vehicle.

Then, the following observation is crucial for computing highly critical test cases. 

Any test case of the form $tc'=((x_{e_0},v_{e_0}),(X'_a[n],V'_a[n]),(X'_f[n],V'_f[n]))$ is less critical than a test case $tc=((x_{e_0},v_{e_0}),(X_a[n],V_a[n]),(X_f[n],V_f[n]))$, if at each time step $i$, $X_{a_i} \leq X'_{a_i}$, $V'_{a_i}\leq V_{a_i}$, and $X_{f_i}\leq X'_{f_i}, V_{f_i} \leq V'_{f_i}$. In fact, any safe policy for $tc$ is also a safe policy for $tc'$.

This result shows that given $tc'$, we can always obtain a more critical test case $ tc$, using constant limit values for the distance between vehicles and the speed functions of the arriving vehicle and the vehicle in front. In fact, we have $V'_a (n)\leq vl$ and $0\leq V_f(n)$. 

So, we can infer that any test case $tc'=((x_{e_0},v_{e_0}),(X'_a (n),V'_a (n)),(X'_f (n),V'_f (n)))$ is less critical than $tc=((x_{e_0},v_{e_0}),(x_{a0} - vl*n, vl),(x_{f0}, 0))$. That is, $tc$ describes scenarios where the arriving vehicle approaches the conflict point at the maximal allowed speed $vl$, and the front vehicle is static. Furthermore, $x_{a0},x_{f0}$ denote the initial distances from the conflict point of the arriving and the front vehicle, respectively. 

These maximal criticality test cases $tc$ are easy to analyse as shown below. We simplify their representation by giving a tuple of parameters, $tc_c =(x_e,v_e,x_a,x_f)$, where $x_e,x_a,x_f$, represent the initial distances of the vehicles from the conflict point and $v_e$ is the initial speed of the ego vehicle. At this point, it is important to note that for the considered adverse scenarios, the ego vehicle has to choose between two different types of policies. 

\begin{itemize}
    \item A \textit{cautious policy}, when the ego vehicle estimates that distances $x_a$ and $x_f$ are too small, and if it moves to cross the conflict point, there is a risk of collision with the arriving or the front vehicle. 
    \item A \textit{progress policy}, when the ego vehicle estimates that by accelerating it will reach the conflict point before the arriving vehicle located at distance $x_a$, while the distance $x_f$ is large enough to brake safely before the front vehicle. 
\end{itemize}

The feasibility of each one of these policies clearly depends on the parameters of $\hat{tc}$ but also on the capabilities of the ego vehicle to accelerate and decelerate, characterized by the following functions: 
\begin{itemize}
    \item The \textit{braking function} $B(v)$ that gives for a given speed $v$ of the ego vehicle, the distance needed to brake safely and stop.
    \item The \textit{acceleration time function} $TA(x,v)$ that gives the time needed to travel distance $x$ from the initial speed $v$.
    \item The \textit{acceleration speed function} $VA(x,v)$ that gives the speed reached when traveling distance $x$ from the initial speed $v$.
\end{itemize}

These functions that we call A/D (acceleration/deceleration) functions are either mathematically defined by autopilots developers or can be estimated experimentally \cite{li2024rigorous}. 

We can now give conditions for the existence of safe test cases relating the parameters of $\hat{tc}$ to A/D functions. 

For safe cautious policies, the ego vehicle should keep its distance $x_e$ such that $B(v_e )\leq x_e$.  

For safe progress policies, two conditions should be satisfied: 
\begin{itemize}
    \item The ego vehicle should have enough time to reach the conflict point before the arriving vehicle, that is $TA(x_e,v_e) \leq x_a/vl$.
    \item The ego vehicle reaching the conflict point at speed $VA(x_e,v_e))$ can safely brake before the front vehicle, that is $B(VA(x_e,v_e)) \leq x_f$.
\end{itemize}

This result shows that for any elementary adverse test case there exist most critical test cases $\hat{tc}$ under the safety condition $x_e\leq B(v_e)$, such that $\hat{tc}=(x_e,v_e,\hat x_a,\hat x_f)$, where $\hat x_a= TA(x_e,v_e)*vl$ and  $\hat x_f = B(VA(x_e,v_e))$. Any other safe progress policy is possible only if $\hat x_a \leq x_a$ and $\hat x_f \leq x_f$.

\section{Autopilot requirements for effective and valid testing}
\label{sec:autopilot-requirements}

Traditional test methods rely on worst-case analysis and coverage criteria that simplify the task by avoiding exhaustive tests that are impossible to perform on complex systems. Based on the framework presented in the previous section, we show that transferring these techniques to ADS requires rational and determinate autopilots. We provide experimental results showing that eight open-loop autopilots fail to satisfy these properties.

\subsection{Autopilot rationality}
\label{sec:rationality}

The assumption that passing a critical test implies passing less critical tests, often adopted for testing real-time systems and hardware systems using worst-case reasoning, is generally not applicable to ADS. It requires that ADS autopilots consistently select control functions based on the criticality of test cases.  Tests carried out on existing autopilots~\cite{li2024rigorous,li2025comprehensive,tian2018deeptest,majumdar2019paracosm} show that, as a general rule, they do not meet this requirement, which prevents the application of worst-case analysis. Such an anomaly can also be related to the property of rationality of agents~\cite{newell1980physical,verschure2003real} stipulating that “If the system wants to attain goal $G$ and knows that to do act $A$ will lead to attaining $G$, then it will do $A$.” This practically means that the autopilot always chooses among the possible control policies the most appropriate. So, if it succeeds for a given test case, it will also be successful for a less critical one. 

Using the framework for calculating critical and safe test cases presented in the previous section, we characterize different types of failures and, in particular, the lack of rationality of elementary adverse scenarios.

\begin{figure}[h]
    \centering
    \includegraphics[width=0.5\linewidth]{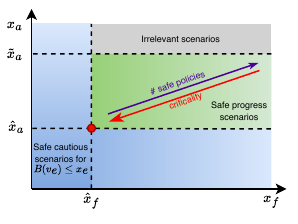}
    \vspace{-15pt}
    \caption{Theoretical decomposition of the test space defined by a most critical test case}\label{fig:generation-theoretical}
\end{figure}

Figure \ref{fig:generation-theoretical} shows a characterization of the space of all possible elementary adverse scenarios parameterized by the values of $x_a$ and $x_f$, for a given position $x_e$ and speed $v_e$ of the ego vehicle. The green zone shows safe progress scenarios that are relevant for $\hat x_a \le x_a \le  \tilde{x}_a$ and $\hat x_f \le x_f$
\begin{itemize}
    \item For values $x_a\le \hat x_a$ or $\hat x_f \le x_f$  there is no safe progress policy. For safety, the autopilot should remain cautious by respecting the constraint $B(v_e) \le x_e$.
    \item For values $\hat x_a \le x_a$ and $\hat x_f \le x_f$ there exist safe policies. For increasing values of $x_a$ and $x_f$, the number of policies increases while their criticality decreases. 
    \item Finally, for values of $\tilde{x}_a \leq x_a$ we have irrelevant test cases for which the ego vehicle can reach the conflict point before the arriving vehicle without modifying its speed, that is, $x_e/v_e \le x_a/v_a * v_l$, which gives $\tilde x_a = (x_e/v_e) v_l$.
\end{itemize}

The most critical test case corresponds to $(\hat x_f, \hat x_a)$ and theoretically passing this test case should imply success for all the other less critical cases. 

\begin{figure}[htbp]
  \centering
  % 第二个子图
  \begin{minipage}[b]{0.49\textwidth}
    \centering
    \includegraphics[width=\linewidth]{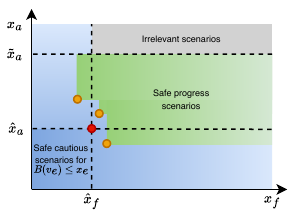}
    \vspace{-20pt}
    \caption{Decomposition obtained in practice}\label{fig:generation-practice}
  \end{minipage}
  \hfill
  \begin{minipage}[b]{0.49\textwidth}
    \centering
    \includegraphics[width=\linewidth]{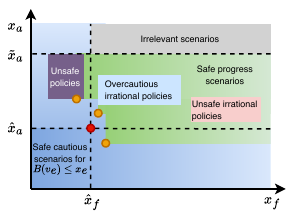}
    \vspace{-20pt}
    \caption{Frequently observed failures}\label{fig:generation-observed}
  \end{minipage}
\end{figure}

Note that in practice, it is hard to estimate precise values of $\hat x_a$ and $\hat x_f$ because the A/D functions are not known exactly or due to other measurement errors. However, we can apply a heuristic approach by exploring test cases defined by pairs $(x_a,x_f)$ around $(\hat x_a, \hat x_f)$. If the test case for some $(x_a,x_f)$ passes, then for rational autopilots, we can infer success for all the test cases $(x_a,x_f) \le (x'_a,x'_f)$. We can thus easily cover all the test space by focusing only on a small number of test cases around $\hat x_a, \hat x_f$. Figure \ref{fig:generation-practice}   illustrates this exploration method where the so identified test cases, denoted by yellow dots, delimit the safe progress zone as an approximation of a convex envelope.  

\begin{table}[h]
    \centering
    \vspace{10pt}
    \caption{The test results for 8 autopilots under 4 scenario types}
    \vspace{-10pt}
    \label{tab:results}
    \adjustbox{max width=\textwidth}{
    \setlength{\tabcolsep}{3pt}
    \begin{tabular}{c|c|c|c|c|c|c|c|c}
    \hline
        \multirow{2}{*}{\textbf{Scenario type}} & \multicolumn{8}{c}{\textbf{Autopilot}} \\
        \cline{2-9}
         & \textbf{Apollo} & \textbf{Autoware} & \textbf{Carla} & \textbf{LGSVL} & \textbf{InterFuser} & \textbf{MILE} & \textbf{Transfuser} & \textbf{LMDrive} \\
         \hline
         \begin{tabular}{c}\textbf{Merge with}\\\textbf{a yield sign}\end{tabular} & 
         \begin{tabular}{c}TF (0.80\%)\\IO (12.7\%) \end{tabular} & 
         \begin{tabular}{c}TF (18.0\%)\end{tabular} & 
         \begin{tabular}{c}TF (9.0\%)\end{tabular} & 
         \begin{tabular}{c}TF (31.5\%)\end{tabular} & 
         \begin{tabular}{c}TF (5.1\%)\\IS (1.7\%) \end{tabular} & 
         \begin{tabular}{c}TF (9.5\%)\end{tabular} & 
         \begin{tabular}{c}TF (6.8\%)\\IS (4.2\%)\end{tabular} & 
         \begin{tabular}{c}TF (2.5\%)\\IS (4.9\%)\end{tabular} \\
         \hline
         \textbf{Lane Change} & 
         \begin{tabular}{c}TF (29.4\%)\\IS (3.9\%) \end{tabular} & 
         \begin{tabular}{c}OF-PD (4/4)\end{tabular} & 
         \begin{tabular}{c}TF (17.5\%)\\ OF-PD (2/4)\end{tabular} & 
         \begin{tabular}{c}TF (10.5\%)\end{tabular} & 
         \begin{tabular}{c}TF (12.5\%)\\IS (3.8\%) \end{tabular} & 
         \begin{tabular}{c}OF-PD (2/2)\end{tabular} & 
         \begin{tabular}{c}OF-PD (2/2)\end{tabular} & 
         \begin{tabular}{c}OF-PD (2/2)\end{tabular} \\
         \hline
         \begin{tabular}{c}\textbf{Intersection with}\\\textbf{yield signs}\end{tabular} & 
         \begin{tabular}{c}TF (55.4\%) \end{tabular} & 
         \begin{tabular}{c}TF (61.2\%)\end{tabular} & 
         \begin{tabular}{c}TF (47.5\%)\end{tabular} & 
         \begin{tabular}{c}TF (58.1\%)\end{tabular} & 
         \begin{tabular}{c}TF (65.5\%)\end{tabular} & 
         \begin{tabular}{c}TF (60.3\%)\end{tabular} & 
         \begin{tabular}{c}TF (71.5\%)\end{tabular} & 
         \begin{tabular}{c}OF-SF (2/2)\end{tabular} \\
         \hline
         \begin{tabular}{c}\textbf{Intersection with}\\\textbf{traffic lights}\end{tabular} & 
         \begin{tabular}{c}TF (27.9\%) \end{tabular} & 
         \begin{tabular}{c}TF (16.7\%)\end{tabular} & 
         \begin{tabular}{c}OF-PD (2/4)\end{tabular} & 
         \begin{tabular}{c}OF-SF (4/4)\end{tabular} & 
         \begin{tabular}{c}TF (6.7\%)\end{tabular} & 
         \begin{tabular}{c}TF (16.0\%)\\IS (7.3\%)\end{tabular} & 
         \begin{tabular}{c}OF-PD (2/2)\end{tabular} & 
         \begin{tabular}{c}OF-SF (2/2)\end{tabular} \\
         \hline
    \end{tabular}
    }
    
    \vspace{2pt}
    \raggedleft
    \hfill \footnotesize{Test results: Transition safety failures (TF), Irrational safety failures (IS), Irrational overcaution (IO), \\ Overall failures involving performance degradation (OF-PD), Overall safety failures (OF-SF)}
\end{table}

The application of this analysis to four modular autopilots, Apollo, Autoware, Carla, and LGSVL, as well as to four end-to-end AI autopilots, InterFuser, MILE, Transfuser, and LMDrive, reveals mainly four different types of property violations shown in Table \ref{tab:results}. The results are presented for four scenario types, including merge with a yield sign, lane change, intersection with yield signs, and intersection with traffic lights. 

The failure types are labeled with their frequencies relative to the number of test cases, except for overall failures, which occur for an initial state of the ego vehicle for all combinations of the values of $x_a$ and $x_f$. For these failures, the frequency is reported as the number of such initial states relative to the total number of initial states considered.

Figure \ref{fig:generation-observed} illustrates some of the most frequently observed failures.

The most frequent failures are \textit{transition safety failures} (TF) occurring in the transition from cautious to progress policies. It often happens that the autopilot overestimates the capabilities of the ego vehicle to reach the conflict point before the arriving vehicle (purple zone). This is probably due to the fact that such a transition corresponds to a rare situation that requires very precise calculation of the relationships between the distances and speeds of the vehicles involved. An ad hoc autopilot, especially one with AI components, cannot handle such cases safely. As shown in Table \ref{tab:results}, all eight autopilots exhibit this type of failure in at least two scenario types. The highest frequency is observed for Transfuser in the crossing with yield signs scenario, reaching up to 71.5\%.

Another category of failures is \textit{irrational safety failures} (IS) appearing for test cases that are not among the most critical, as indicated in the red zone. After the test case for $(x_a, x_f)$ passed successfully, the autopilot fails for less critical cases $(x'_a,x'_f)$, $(x_a,x_f) \le (x'_a, x'_f)$. As explained, this anomaly can be interpreted as a lack of rationality. The high number of failures observed for some autopilots renders the principle of critical testing completely ineffective. These failures are more pernicious than transition safety failures, which are borderline cases, because they can occur in completely unexpected situations. As shown in Table \ref{tab:results}, end-to-end autopilots experience many more failures of this kind than modular autopilots. Among the four modular autopilots, only Apollo shows such issues with a frequency of 3.9\% during lane changes. In contrast, all four AI autopilots exhibit irrational safety failures in one or two scenario types with frequencies ranging from 1.7\% to 7.3\%.

The third category of failures is \textit{irrational overcaution} (IO), concerning the deterioration in performance indicated by the part of the blue zone that falls within the green zone. These are cautious and safe policies when all the conditions are right for progress. They are therefore irrational overcaution policies, which can cause traffic problems in practice and are rarely addressed in ADS testing work. Among the eight autopilots tested, only the Apollo autopilot exhibited this kind of problem with a high frequency of 12.7\%, when merging with a yield sign.

A fourth category of failures is \textit{overall failures} (OF) observed for all values $(x_a,x_f)$, for a given initial state of the ego vehicle $(x_e,v_e)$, meaning that the autopilot cannot successfully handle a specific traffic pattern, whether it is critical or not. 

We have observed overall failures involving performance degradation (OF-PD), where the ego vehicle does not progress for all values $(x_a,x_f)>(\hat x_a,\hat x_f)$. These failures mainly occur during lane changes, where three of the eight autopilots, Autoware, MILE, and Transfuser, fail to move. Transfuser and Carla autopilots also showed this problem at intersections with traffic lights. 

Another type of overall failure involves inadvertent route changes (OF-RC), where the ego vehicle does not follow the specified route and instead takes a different path when crossing an intersection, which is independent of the front vehicle and the arriving vehicle. This happens with LMDrive at intersections with yield signs or traffic lights.	

There are also overall failures related to safety, where accidents or property violations occur. This type of failure is observed in the LGSVL autopilot at intersections with traffic lights, occurring in all four tested initial states, and in LMDrive at intersections with yield signs, occurring in one of the two tested initial states.

It should be noted that the proposed test method uses the optimal criticality order to systematically explore possible failures for values of $x_a$ and $x_f$ in specific ranges. The critical values $(\hat x_a, \hat x_f)$ provide only guidance on how to explore the test space. However, even without knowing these values, the method requires consistency between the control policies observed when the constraints imposed by the ego vehicle environment are relaxed. From this point of view, the method presented here has similarities with metamorphic testing \cite{li2023simulation}, which uses a similarity relationship between test cases and requires that the corresponding scenarios are not distinguishable for the tested properties.

Note that the lack of rationality can also be observed by checking the consistency of the decisions made by an autopilot for two equally critical test cases. These test cases are logically equivalent, as they allow for exactly the same set of safe control policies. It has been demonstrated that autopilots can pass one test case and fail another simply by rotating the configurations of the vehicles on a map, without changing their relative distances and speeds~\cite{li2024rigorous}.

Other concepts of logical equivalence can be defined relating a test case with a test case that is obtained by adding vehicles whose presence should not logically affect the control policy of the autopilot. For example, given a test case involving a front vehicle and an arriving vehicle, an equivalent test case can be obtained by adding vehicles located behind the arriving vehicle or after the front vehicle. The presence of these vehicles should not logically affect the behavior of the ego vehicle. However, autopilots, particularly those based on AI techniques, can be sensitive to these changes. For instance, we tested a merge scenario where the ego vehicle, equipped with the LMDrive autopilot, starts at the entrance of the merge. With an arriving vehicle 35 meters from the merge point, the ego vehicle successfully merges before the arriving vehicle. However, when a second arriving vehicle is added at 45 meters from the merge point, the ego vehicle becomes cautious and waits before the merge point until the arriving vehicles pass.

\subsection{Autopilot determinacy}
\label{sec:determinacy}

Traditional methods rely on coverage criteria that cannot be transferred to ADS without assuming the reproducibility of the observed behavior. The concept of coverage relies on the assumption that the success of a test case generated from an initial state $q_0$, a sequence of states  $q_0,...,q_i,...,q_n$, implies the success of test cases from any visited state $q_i$. The objective is then to generate sequences of states that explore the entire state space of an abstract model of the system under test.

Applying this principle to ADS testing raises several questions, which we will attempt to analyze below. Unlike traditional hardware or software testing, which focuses on systems characterized by a single transition relation, an autopilot can apply various control policies depending on the context and status of autonomous vehicles and the states of the surrounding obstacles.

A policy function $pol$ is a function $pol: Q\times S^n \to Q^n$, which from a given test case $tc=(q_0,s_0,...,s_i,...,s_n)$, specified by an initial state $q_0$ of the ego vehicle and a sequence of environment states, produces a corresponding control policy described by a sequence of system states. More precisely, $pol(q_0,s_0,...,s_i,...,s_n )=q_0,...,q_i,...,q_n$.

Clearly, for the test case  $tc$ and the resulting control policy, $q_0,...,q_i,...,q_n$, it is possible to define test cases $tc_i=(q_i,s_i,...,s_n)$ from the states visited with the corresponding environment behavior.

A question naturally arises: if the autopilot is tested for a test case $tc_i=(q_i,s_i,...,s_n)$ and applies a control function $pol_i$, does it generate a control policy $pol_i(tc_i )=q_i,...,q_n$, which is consistent with the control policy generated from $tc$. If so, this means that validating the autopilot for $tc$ implies validation for $tc_i$ and greatly simplifies the testing problem. Otherwise, passing $tc$ does not necessarily imply validation for the test cases $tc_i$ so defined.  

An autopilot is \textit{determinate} for a policy function $pol$ applied to a test case $tc=(q_0,s_0,...,s_i,...,s_n)$ such that $pol(q_0,s_0, ..., s_i,...,s_n )=q_0,...,q_i,...,q_n$, if the control functions $pol_i$ applied to cases $tc_i=(q_i,s_i,...,s_n)$ generate consistent control policies, $pol_i(tc_i)=q_i,...,q_n$. We say that the autopilot is determinate if it is determinate for any policy function it applies.

It should be noted that determinacy differs from determinism, which, in a given situation, implies a single outcome and excludes any possibility of choice, as in classical physics. On the contrary, determinacy allows the autopilot to choose between possible policy functions for a given test case, but requires consistency in the control policies generated for the considered test case.

In practice, we can consider a driving scenario as a sequence of maneuvers for which the autopilot applies successfully corresponding policy functions $pol_1,...,pol_k$ for which it is determinate. To achieve determinacy for a given policy function, it is enough to consider that it is generated by iterative application of a transition function $\pi$, in the following manner.

Consider a function  $\pi:Q\times S\to Q$, where $\pi(q,s)=q'$ means that the autopilot moves from state $q$ to state $q'$ when the environment state is $s$. The function $\pi$ can be naturally extended for sequences of states of the environment: $\pi(q,sS)= \pi(\pi(q,s),S)$ and $\pi(q_0,\epsilon)=q$, where $\pi(q,s)$ is the state reached from $q$ when the environment state is $s,S$ is a sequence of environment states, and $\epsilon$ is the empty sequence. Clearly, $\pi(q_0,s_0,...,s_k)=q_k$ can be computed by successive application of $\pi(q_i,s_i)=q_{i+1}$, for $i=0,...,(k-1)$.  

Using a transition function $\pi$, an autopilot can generate determinate control policies by application of the rules: $pol(q,sS) =q\ pol(\pi(q,s),S)$ and $pol(q,\epsilon)=q$, where $S$ is a sequence of environment states and $\epsilon$ is the empty sequence. From this definition, it follows the following composability property: 
$pol(q_0,s_0,...,s_i,...,s_n )=pol(q_0,s_0,...,s_{i-1})\ pol(\pi(q_0,s_0,...,s_{i-1}), s_i,..., s_n)$

This means that the generated control policy from state $q_0$ by applying a sequence of $n$ environment states is the concatenation of the control policies resulting from application of the sequence of the first $i$ states, and then the application of the sequence of $(n-i+1)$ states, from the state reached after application of the first sequence.

The functional decomposition of the autonomous driving function into simple control policies greatly simplifies the problem of testing autopilots \cite{corso2020scalable,yang2024roadgen,amersbach2019functional}. However, we will not examine this general problem and will instead focus on the conditions under which the determinacy of simple deceleration or acceleration policies can be achieved, given that any control policy can be decomposed into sequences of such elementary policies, possibly interspersed with constant-speed policies. Their determinacy therefore appears to be a necessary condition for the determinacy of composite policies.

\begin{figure}[h]
    \centering
    \includegraphics[width=0.9\linewidth]{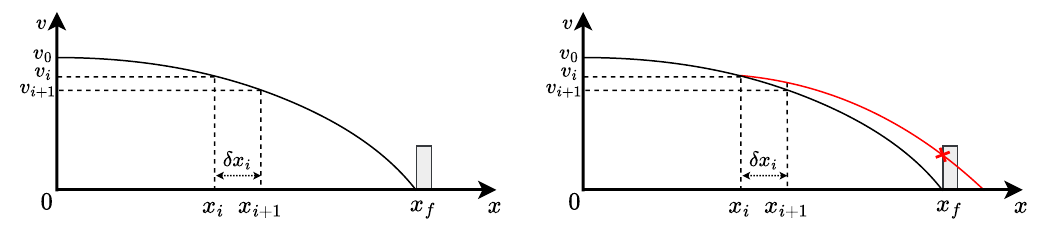}
    \vspace{-15pt}
    \caption{Determinate and non-determinate braking policy }
    \vspace{-5pt}
    \label{fig:braking-policy}
\end{figure}

Consider the simple maneuver of the ego vehicle braking before a static object at a distance $x_f$. The control policy can be described by a sequence of decreasing speeds and increasing traveled distances $(v_0,x_0 ),...,(v_n,x_n )$, such that $x_0=0,x_n=x_f$ (Figure \ref{fig:braking-policy}). The question is whether testing that the speed of the ego will be zero before the obstacle from the state $(v_0,x_0)$, validates the behavior of the vehicle for all the visited states $(v_0,x_0 ),...,(v_n,x_n)$. 

Assume that a control policy is generated by a speed braking function $VB(v,x)=v'$ that gives the speed $v'$ reached from speed $v$ by braking for distance $x$. This function determines the braking function $B(v)=x$ used in Section \ref{sec:rigorous-framework}: $B(v)=x$ if $VB(v,x)=0$.

Then, the application of the braking function from $(x_0,v_0)$ for distances $x_i$ and $x_{i+1}$, gives speeds $VB(v_0,x_i)=v_i$ and $VB(v_0,x_{i}+\delta x_i)=v_{i+1}$, where $x_{i+1}=x_i +\delta x_i$,  respectively. For determinacy, the policy applied from $(x_i,v_i)$ for distance $\delta x_i$ should be equal to $v_{i+1}$, that is $VB(v_i,\delta x_i )=v_{i+1}$. That is, $VB(v_0,x_i + \delta x_i) = VB(VB(v_0,x_i), \delta x_i)$.

Note that the determinacy requirement is trivially satisfied when $VB$ is defined for a constant braking rate $b$. We have $VB(v,x)=v'$ if $(v'^2-v^2)/2b=x$. Using this definition, we get for $VB(v_0,x_i)= v_i$, the equation $(v_0^2-v_i^2)/2b=x_i$, and for $VB(v_0, x_{i+1}) =v_{i+1}$, the equation $(v_0^2-v_{i+1}^2)/2b=x_{i+1}$. From which we obtain the equation $(v_{i+1}^2-v_i^2)/2b=x_{i+1}-x_i$ which holds if $VB(v_i,x_{i+1} -x_i) = v_{i+1}$.

\begin{figure}[h!]
  \centering
  \begin{subfigure}[b]{0.48\textwidth}  % 宽度可根据需要调整
    \includegraphics[width=\linewidth]{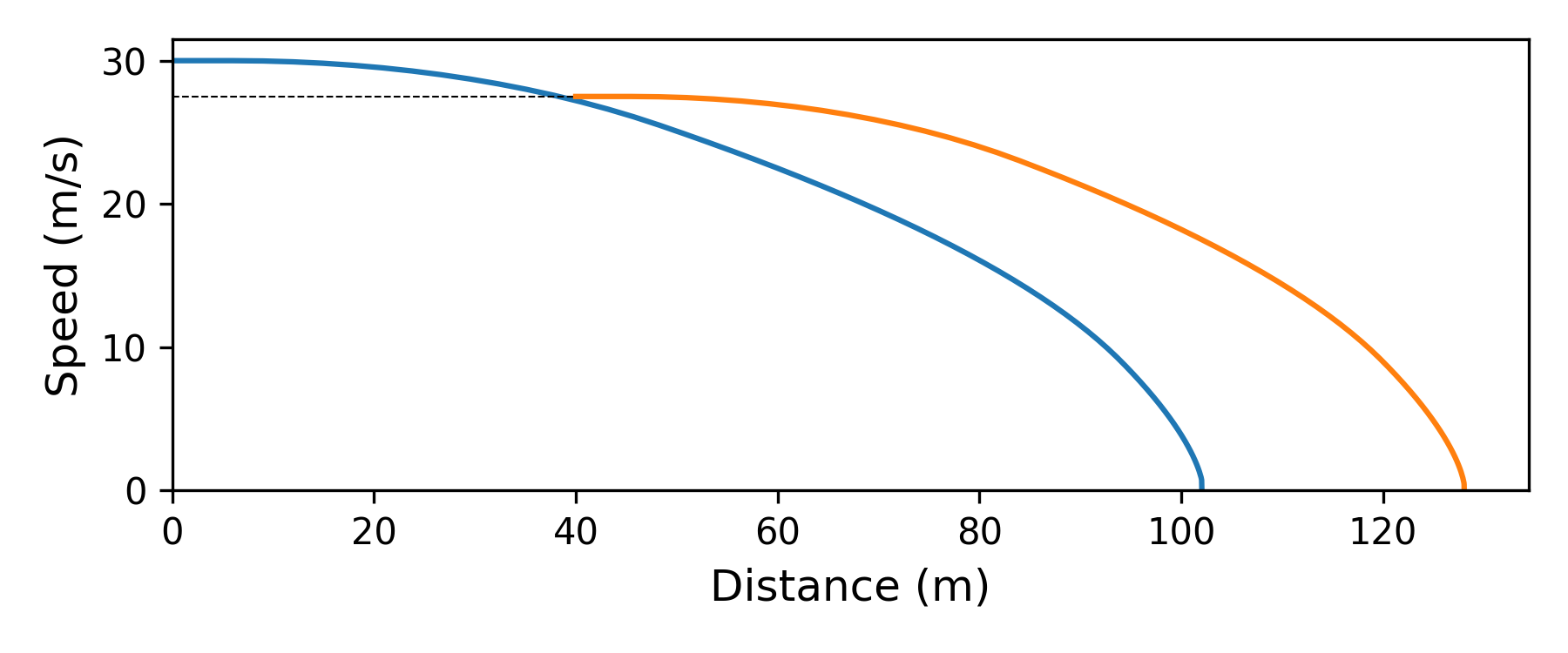}
    \caption{Apollo}
    \label{fig:sub1}
  \end{subfigure}
  \hfill
  \begin{subfigure}[b]{0.48\textwidth}
    \includegraphics[width=\linewidth]{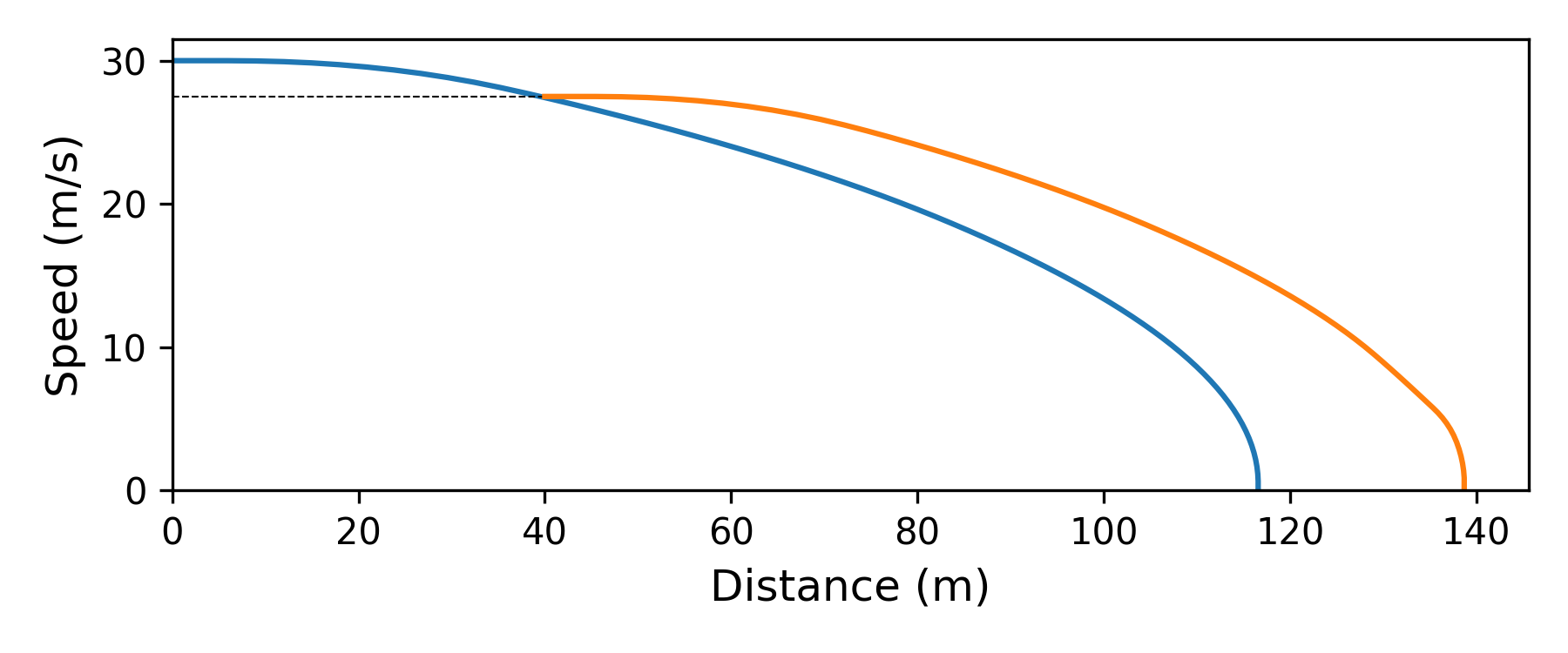}
    \caption{Autoware}
    \label{fig:sub2}
  \end{subfigure}
  \vspace{-5pt}
  \caption{Braking curves for Apollo and Autoware autopilots }
  \label{fig:braking-exp}
\end{figure}

We observed that the determinacy requirement for braking functions is not satisfied by seven of the eight tested autopilots, with the exception of the LGSVL autopilot. For these autopilots, if the vehicle brakes safely from $(x_0, v_0)$, it may not brake safely from $(x_i, v_i)$, which is a state of the braking curve. Thus, without determinacy, even the verification of very simple safety requirements, such as braking before an obstacle, may require an infinite number of tests. Figure \ref{fig:braking-exp} shows the braking curves for Apollo and Autoware, illustrating the minimum required distances when braking from 30 m/s and 27.5 m/s. For both autopilots, braking from 27.5 m/s requires a much longer distance than braking from the same distance, after decelerating from 30 m/s. The differences in braking distance are 26.0 meters for Apollo and 22.1 meters for Autoware, respectively.

\begin{figure}[h]
    \centering
    \includegraphics[width=0.45\linewidth]{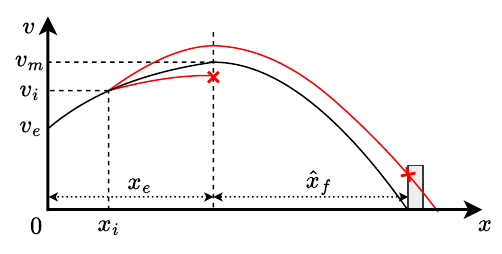}
    \vspace{-15pt}
    \caption{Determinate and non-determinate progress policy}
    \label{fig:progress-policy}
\end{figure}

The same type of reasoning can be applied to acceleration functions. Consider the scenario of a progress maneuver of the ego vehicle starting with speed $v_e$ at a distance $x_e$ from a critical zone where it can have a collision with an arriving vehicle at distance $\hat x_a$  and a front vehicle stopped on its route at distance $\hat x_f$, where $\hat x_a, \hat x_f$ are the critical values for avoiding collision (Figure \ref{fig:progress-policy}). If the test case $\hat {tc} = (x_e,v_e,\hat x_a,\hat x_f)$ passes when the ego vehicle's autopilot applies a policy function $pol$ and is determinate, then it will pass the test cases starting from any state $(x_i,v_i)$ of the generated control policy $pol(\hat {tc}) = (x_0,v_0),...,(x_i, v_i),...,(x_n, v_n)$, with state $(x_0,v_0 )=(0,v_e)$ and state $(x_e, v_m)$ at the conflict point. In fact, determinacy for $VA(v,x)$ implies in particular that the speed reached from a point $(x_i,v_i)$ by traveling distance $x_e-x_i$ will be equal to the speed reached from $(x_0,v_e)$ by traveling $x_e$. That is, $VA(v_e,x_e )=VA(VA(v_e,x_i), (x_e-x_i))$. Any possible deviation of the generated control policy from state $(x_k, v_k)$, resulting in a speed at $x_e$ different from $v_m$, would cause the test case to fail, as shown by the red curves of Figure \ref{fig:progress-policy}. If the acceleration policy starting from $(x_i, v_i)$ results in a speed higher than $v_m$ at $x_e$, the ego vehicle will be unable to stop safely before reaching the front vehicle. Conversely, if it results in a speed lower than $v_e$, the ego vehicle cannot pass the critical zone in time and may collide with the arriving vehicle.

\begin{figure}[h]
    \centering
    \includegraphics[width=0.5\linewidth]{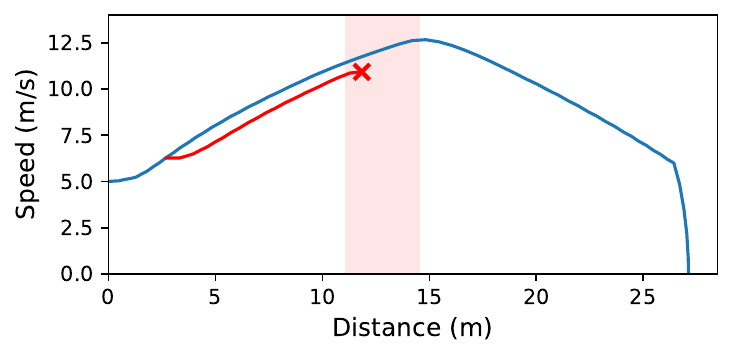}
    \vspace{-10pt}
    \caption{Non-determinate progress policy for the Carla autopilot at an intersection with yield signs}
    \label{fig:acc-carla}
\end{figure}

A check of the acceleration policies for the eight open autopilots shows that they are not determinate. Figure \ref{fig:acc-carla} illustrates an example of a non-determinate progress policy exhibited by the Carla autopilot at an intersection with yield signs. When the ego vehicle starts at $v_e=5.0$ m/s and $x_e=11.05$ m, it successfully handles a critical test case, where $\hat{x}_a = 27.6$ m and $\hat{x}_f = 12.6$ m, by applying the policy indicated in blue curve, accelerating to pass through the critical zone (red region), and braking before reaching the front vehicle. In this case, the safe progress policy passes through a state with $(x_i, v_i) = (2.7,\ 6.3)$ and $x_a = 20.75$. However, when the ego vehicle starts from $(x_i, v_i) = (2.7,\ 6.3)$ with $x_a = 20.75$, it applies the policy as indicated in red that has a lower acceleration rate and collides with the arriving vehicle within the critical zone, indicated by the red cross in Figure \ref{fig:acc-carla}.

\subsection{Improving coverage by test case partitioning}
\label{sec:monotonicity}

Traditional methods use test case partitioning techniques that allow test cases to be grouped into classes, so that only one condition needs to be tested to determine the success or failure of all items in the class. We show that if A/D functions satisfy simple monotonicity conditions, it is possible to apply a partitioning technique that can greatly simplify the exploration of the test space.

\begin{figure}[h]
    \centering
    \begin{minipage}{0.48\linewidth}
        \includegraphics[width=\linewidth]{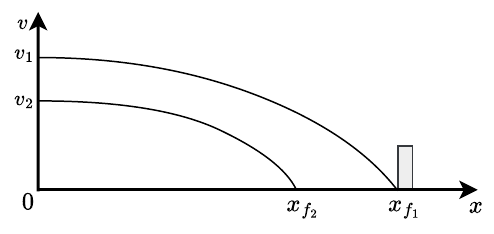}
        \vspace{-19pt}
        \caption{Monotonicity of braking functions}
        \label{fig:mono-brake}
    \end{minipage}
    \begin{minipage}{0.48\linewidth}
    \includegraphics[width=\linewidth]{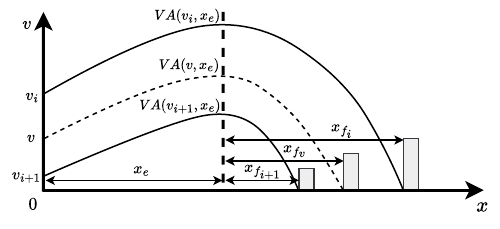}
    \vspace{-15pt}
    \caption{Monotonicity for acceleration functions}
    \label{fig:mono-acc}
    \end{minipage}
\end{figure}

First, let us assume that $VB$ is monotonic with respect to $v$: $VB(v,x)\ge VB(v',x)$ if $v\ge v'$. Let us return to the basic scenario of braking in front of an obstacle located at a distance $x_f$ (Figure \ref{fig:mono-brake}). If $VB$ is monotonic, safe braking from $v$ such that $VB(v,x_f)=0$ implies safety from any speed $v'\le v$ for an obstacle located at $x_f$.

Now, let us consider progress policies for a distance $x_e$ of the ego vehicle before the conflict point and a distance $x_f$ of the front vehicle after the conflict point. Let $v_0$ be the maximum permitted speed of the ego vehicle from distance $x_f$. The results of section \ref{sec:optimal-criticality} allow us to find the minimum distances $\hat{x}_a$ and $\hat{x}_{f}$ for which there are safe policies for this configuration. 

Assume that in addition to the monotonicity of $VB$, the following properties are satisfied by $TA$ and $VA$:	 
$TA(v,x)\le TA(v', x)$ if $v'\le v$ and $VA(v,x)\le VA(v', x)$ if $v\le v'$.

Instead of testing from the very large number of possible speeds $v\le v_0$, we can test for speed values varying in intervals defined by a decreasing sequence of speeds $v_0 > v_1 > v_2 > ... > v_n$. We provide sufficient conditions for the existence of safe policies for speeds $v$ in any interval $v_{i+1}< v <v_i$, which limits the number of test cases to the number of intervals (Figure \ref{fig:mono-acc}).

Let $\hat x_{a_i}$ be the least distance of the arriving vehicle to avoid collision from $v_i$, and $\hat x_{f_{i+1}}$ be the least distance of the front vehicle to avoid collision from $v_{i+1}$. Then, we can show that under the monotonicity assumption, there exist safe policies for speeds $v$, $v_i\le v \le v_{i+1}$ for $x_a$, $\hat x_{a_{i+1}}\le x_a$ and for $x_f$, $\hat x_{f_{i}}\le x_f$, respectively. 

For initial speeds $v_i$ and $v_{i+1}$, the ego vehicle will reach the conflict point with respective speeds $VA(v_i,x_e)$ and $VA(v_{i+1}, x_e)$ such that $VA(v_{i+1},x_e)< VA(v_i,x_e)$ and travel times $TA(v_i,x_e)$ and $TA(v_{i+1},x_e)$ such that $TA(v_i,x_e)< VA(v_{i+1},x_e)$. Then, if the initial speed of the ego vehicle is $v$, $v_{i+1}\le v \le v_{i}$, the collision with the arriving vehicle will be avoided for all $x_a$, such that $\hat x_{a_{i+1}} \le x_a$ with $TA(v_{i+1},x_e)*vl=\hat x_{a_{i+1}}$. Furthermore, for a speed $v$, $v_{i+1}\le v\le v_i$, the collision will be avoided for all $x_f$ such that $\hat x_{f_0} \le x_f$ with $B(VA(v_0,x_e)) = \hat x_{f_i}$.  

Thus, there exist safe test policies for test cases with initial speed in the interval $[v_{i+1},v_i]$, for values $x_a$ and $x_f$ such that $\hat x_{a_{i+1}} \le x_a$ and $\hat x_{f_i} \le x_f$, respectively.

\begin{figure}[h]
    \centering
    \includegraphics[width=0.45\linewidth]{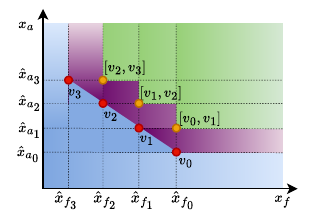}
    \vspace{-10pt}
    \caption{Decomposition of the test space defined by safe policies of a speed partition}
    \label{fig:test-space-decomposition}
\end{figure}

In Figure \ref{fig:test-space-decomposition}, we reproduce the decomposition of the test space of Figure \ref{fig:generation-theoretical}, with the red points showing the most critical test cases for given initial speeds, $v_0,v_1,v_2,v_3$. The orange points show constraints on $x_a$ and $x_f$ for which test cases in the corresponding intervals of the considered partitioning have safe policies if the ego autopilot is rational. The blue area covers test cases for which only cautious policies are possible. The purple area covers test cases for which there exist safe policies that cannot be covered by the considered approximation. It is delimited by an envelope generated by the most critical test cases when the initial speed decreases continuously from $v_0$. We can thus compute the coverage achieved as the ratio: $green\_area/(green\_area+purple\_area)$ for rational and determinate autopilots. This area can be reduced if finer partitions of the speed range are considered. 

\subsection{A test methodology}

We propose a methodology that, for rational and determinate autopilots, significantly improves the effectiveness of autopilot testing for the elementary adverse scenarios considered. In fact, test cases are characterized by four independent parameters, $x_e,v_e,x_a,x_f$, which are the distance and speed of the ego vehicle from the conflict point, and the distances of the arriving and the front vehicle from this point. Focusing on critical situations, we consider, as previously, the speed of the arriving vehicle equal to the maximum allowed speed $vl$ and the speed of the front vehicle equal to zero. 

Autopilot rationality allows inferring for given $x_e, v_e$, from the success of $\hat {tc} = ( x_e, v_e, \hat x_a, \hat x_f,)$, with $\hat x_a= TA(x_e,v_e )* vl$ and $\hat x_f  = B(VA(x_e,v_e ))$,  the success of all the test cases $tc_c =(x_e,v_e,x_a,x_f )$, such that $\hat x_a  \le x_a, \hat x_f \le x_f$. These are exactly all the test cases for which safe progress policies exist. Even though the theoretical values of $\hat x_a$ and $\hat x_f$ cannot be calculated precisely, the experimental results show that their estimated values are sufficiently accurate to identify a set of maximally critical test cases for which safe progress policies exist, covering all other test cases. 

Thus, rationality makes it possible to reduce the validation of a very large number of test cases for a given pair $(x_e, v_e)$, to the validation of a limited number of test cases.

On the other hand, determinacy allows to infer from success of $(x_e, v_e, x_a, x_f)$, success for all subsequent states  $(x_{e_i},v_{e_i},x_{a_i},x_{f_i})$ visited over the execution of the same policy. Thus, for determinate autopilots, it is enough to consider test cases for the maximal distance $x_{e_{max}}$ from which a policy is applied. These test cases for rational autopilots are of the form $(x_{e_{max}},v_{e_i},\hat x_{a_i},x_{f_i})$ for different $v_{e_i}$, given that $\hat x_{a_i}, \hat x_{f_i}$ are computed from $x_{e_{max}},v_{e_i}$. 

Finally, the monotonicity of A/D functions allows to cope with the complexity of exploring all the test cases $( x_{e_{max}},v_{e_i}, \hat x_{a_i},\hat x_{f_i})$ for different values of $v_{e_i}$. The success of two test cases $(x_{e_{max}},v_{e_i},\hat x_{a_i}, \hat x_{f_i})$, and $(x_{e_{max}},v_{e_{i+1}},\hat x_{a_{i+1}},\hat x_{f_{i+1}})$, for speeds $v_{e_i}$ and $v_{e_{i+1}}$, implies success for any $(x_{e_{max}}, v, \hat x_{a_v},\hat x_{f_v})$, such that $v_{e_{i+1}}\le v\le v_{e_i}$. Therefore, by partitioning the initial speed domain and starting from distance $x_{e_{max}}$, we can cover a very large number of test cases, with the degree of coverage depending on the granularity of the partition.

\begin{figure}[h]
    \centering
    \includegraphics[width=\linewidth]{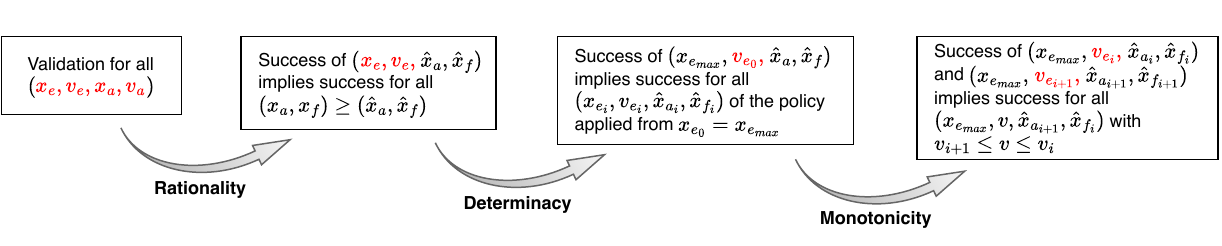}
    \vspace{-25pt}
    \caption{Test methodology}
    \label{fig:test-methodology}
\end{figure}

Figure \ref{fig:test-methodology} illustrates the steps of the test methodology. At each step, based on the corresponding property of the autopilot, the number of test cases can be reduced by considering only the variables highlighted in red within their respective domains.

\section{Conclusion}
\label{sec:conclusion}

This paper provides a critical analysis of testing approaches used to validate the essential properties of ADS. As these are critical systems, it is important to provide irrefutable safety guarantees first and foremost, and to a lesser extent, performance guarantees. We have demonstrated that applying common-sense engineering criteria, such as worst-case analysis and coverage, is not guaranteed to work and requires careful design of autopilots. 

Taking into account the operating conditions of ADS is not as simple as for traditional systems, where it is sufficient for test cases to restrict the definition of the input variables to their domains. For ADS, test cases must be defined according to the evolution of the vehicle, considering its kinetic state and associated control mechanisms. In particular, if tests are to be carried out in extreme situations, care must be taken to ensure that these situations can always be controlled in accordance with the nominal capabilities of the autopilots.

An analysis of the various coverage criteria applied shows that quantitative criteria in terms of distance or time traveled are often not technically tenable, as they do not specify the extent to which the test conditions cover real-life situations. We need generic test methods based on robust validity indicators to ensure that their results reflect real-world outcomes.

Given the nature of ADS, most works are devoted to generating critical tests. However, very few apply criteria that take into account nominal operating conditions. We propose an optimal criticality order and show how it can be calculated for elementary adverse scenarios that cover most conflict situations between the ego vehicle and its environment. 

Critical test methods are based on the assumption that autopilots are rational: validation in the most critical situations implies validation in less critical situations. Our analysis of the test results based on the optimal criticality order shows that five out of eight open autopilots do not meet this assumption. This makes the applicability of critical test methods technically questionable. Rationality is an intrinsic property of autopilots and must be taken into account throughout their design. Although this seems at least theoretically possible in traditional design processes, it is difficult to imagine how rationality can be guaranteed by design for end-to-end AI systems.

A second source of problems lies in the non-determinacy of autopilots, even when they apply the most elementary control functions for braking or acceleration. This calls into question the very foundations of the concept of coverage, since passing a test from an initial state does not guarantee that the test will pass from the states visited. Determinacy allows for overcoming this difficulty. However, the experimental results show that all eight tested autopilots exhibit non-determinacy for basic acceleration policies, and seven of them exhibit non-determinacy for basic braking policies. We show that to achieve the determinacy of these policies, it suffices to implement them by iteratively applying a transition function that can be defined empirically or formally, as is typically the case in dynamic systems and control theory.

We close our analysis by showing how the rationality and determinacy of autopilots can be used profitably to significantly reduce test complexity, particularly when combined with test case partitioning techniques applicable when A/D functions exhibit certain monotonicity properties. This results in a test methodology whose coverage rate can be accurately estimated for the elementary adverse scenarios considered.

We conclude by emphasizing that, as things stand, it seems impossible to obtain the safety guarantees required for ADS without autopilots being designed to be testable, in particular, rational and determinate. Meeting these properties seems to be the only realistic option for providing reliability guarantees for ADS, as ADS are critical systems for which the required level of reliability is in the order of $10^{-8}$ failures per hour of operation. Achieving such guarantees appears infeasible due to the shortcomings of existing test methods, further compounded by problems related to poor autopilot design.
 
\bibliographystyle{ACM-Reference-Format}
\bibliography{references}

\end{document}